\documentclass[superscriptaddress,showpacs,preprintnumbers,amsmath,amssymb,floatfix]{revtex4}
\usepackage{graphicx}
\usepackage{epsfig}
\begin{document}
\title{PHYSICS of TRANSPORT and TRAFFIC PHENOMENA in BIOLOGY:\\
from molecular motors and cells to organisms }
\author{Debashish Chowdhury{\footnote{E-mail: debch@iitk.ac.in 
(Corresponding author)}}}
\affiliation{Department of Physics, Indian Institute of Technology,
Kanpur 208016, India.}
\author{Andreas Schadschneider{\footnote{E-mail: as@thp.uni-koeln.de}}}
\affiliation{Institute for Theoretical Physics, University of Cologne, 
D-50937 K\"oln, Germany.}
\author{Katsuhiro Nishinari{\footnote{E-mail: tknishi@mail.ecc.u-tokyo.ac.jp}}}
\affiliation{Department of Aeronautics and Astronautics,
Faculty of Engineering, University of Tokyo,
Hongo, Bunkyo-ku, Tokyo 113-8656, Japan.}
\date{\today}%
\begin{abstract}
Traffic-like collective movements are observed at almost all levels 
of biological systems. Molecular motor proteins like, for example, 
kinesin and dynein, which are the vehicles of almost all 
intra-cellular transport in eukayotic cells, sometimes encounter 
traffic jam that manifests as a disease of the organism. Similarly, 
traffic jam of collagenase MMP-1, which moves on the collagen fibrils 
of the extracellular matrix of vertebrates, has also been observed 
in recent experiments. Novel efforts have been made to utilize some 
uni-cellular organisms as ``micro-transporters''. Traffic-like 
movements of social insects like ants and termites on trails are, 
perhaps, more familiar in our everyday life. Experimental, theoretical 
and computational investigations in the last few years have led to a 
deeper understanding of the generic or common physical principles 
involved in these phenomena. In this review we critically examine 
the current status of our understanding, expose the limitations of the 
existing methods, mention open challenging questions and speculate on 
the possible future directions of research in this interdisciplinary 
area where physics meets not only chemistry and biology but also 
(nano-)technology. 
\end{abstract}
\maketitle

\section{Introduction} 

Motility is the hallmark of life. From intracellular molecular transport 
and crawling of amoebae to the swimming of fish and flight of birds, 
movement is one of life's central attributes. All these ''motile'' 
elements generate the forces required for their movements by 
{\it actively} converting some other forms of energy into mechanical 
energy. However, in this review we are interested in a special type 
of collective movement of these motile elements. What distinguishes 
a {\it traffic-like} movement from all other forms of movements is 
that traffic flow takes place on {\it ``tracks''} and {\it ``trails''} 
(like those for trains and street cars or like roads and highways for 
motor vehicles) for the movement of the motile elements. From now 
onwards, the term ``element'' will mean the motile element under 
consideration. 

\begin{figure}[ht]
\begin{center}
\includegraphics[angle=-90,width=0.48\columnwidth]{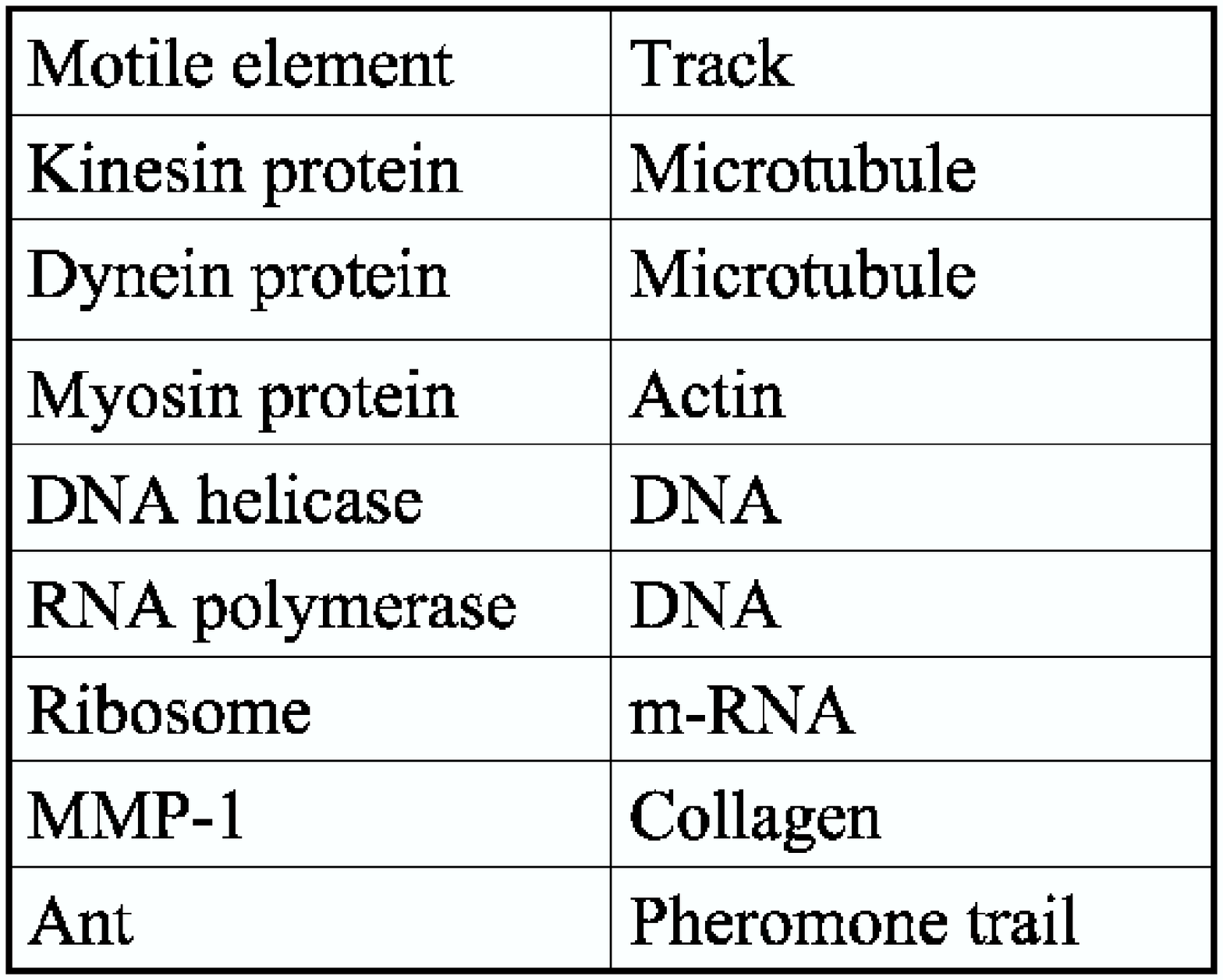}
\end{center}
\caption{Examples of motile elements and the corresponding tracks.
}
\label{fig-table}
\end{figure}

We are mainly interested in the {\it general principles} and common 
trends seen in  the mathematical modeling of collective traffic-like 
movements at different levels of biological organization. We begin at 
the lowest level, starting with intracellular biomolecular motor 
traffic on filamentary rails and end our review by discussing the 
collective movements of social insects (like, for example, ants and 
termites) and vertebrates on trails. Some examples of motile elements 
and the corresponding tracks are shown in fig.\ref{fig-table}.
                                                                               
\subsection{Different types of traffic in biology} 

Now we shall give a few examples of the traffic-like collective phenomena 
in biology to emphasize some dynamical features of the tracks which makes 
biological traffic phenomena more exotic as compared to vehicular traffic. 
In any modern society, the most common traffic phenomenon is that of 
vehicular traffic. The changes in the roads and highway networks take 
place over periods of years (depending on the availability of funds) 
whereas a vehicle takes a maximum of a few hours for a single journey. 
Therefore, for all practical purposes, the roads can be taken to be 
independent of time while studying the flow of vehicular traffic.  
In sharp contrast, the tracks and trails, which are the biological 
analogs of roads, can have nontrivial dependence on time during the 
typical travel time of the motile elements. We give a few examples of 
such traffic. 

\noindent $\bullet$ {\it Time-dependent track whose length and shape 
can be affected by the motile element:} Microtubules, a class of 
filamentary proteins, serve as tracks for two superfamilies of motor 
proteins called kinesins and dyneins \cite{howard,schliwa,hackney}. 
Interestingly, microtubules are known to exhibit an unusual 
polymerization-depolymerization dynamics even in the absence of motor 
proteins. Moreover, in some circumstances, the motor proteins interact 
with the microtubule tracks so as to influence their length as well as 
shape; one such situation arises during cell division (the process is 
called {\it mitosis}). 

\noindent $\bullet$ {\it Time-dependent track/trail created and maintained 
by the motile element:} A DNA helicase \cite{patel,crampton} unwinds 
a double-stranded DNA and uses one of the single strands thus opened 
as the track for its own translocation. Ants are known to create the 
trails by dropping a chemical which is generically called {\it pheromone} 
\cite{wilson}. Since the pheromone gradually evaporates, the ants keep 
reinforcing the trail in order to maintain the trail networks. 

\noindent $\bullet$ {\it Time-dependent track destroyed by the motile 
element:} A class of enzymes, called MMP-1, degrades their tracks 
formed by collagen fibrils \cite{nagase,whittaker}. 

Our aim is to present a critical overview of the common trends in 
the mathematical modelling of these traffic-like phenomena. 
Although the choice of the physical examples and modelling strategies 
are biased by our own works and experiences, we put these in a 
broader perspective by relating these with works of other research 
groups. 
 
This review is organized as follows: the general physical principles 
and the methods of modelling traffic-like collective phenomena are 
discussed in sections \ref{sec-approach}-\ref{sec-intra} while specific 
examples are presented in the remaining sections. 
A summary of 
the various theoretical approaches followed so far in given in section 
\ref{sec-approach}. 
The totally asymmetric simple exclusion process (TASEP), which lies at  
the foundation of the theoretical formalism that we have used successfully 
in most of our own works so far, has been described separately in section 
\ref{sec-asep}.  The Brownian ratchet 
mechanism, an idealized generic mechanism of directed, albeit noisy, 
movement of single molecular motors, is explained in section \ref{sec-single}. 
Traffic of ribosomes, a class of nucleotide-based motors, is considered 
in section \ref{sub-protein}. Intracellular traffic of cytoskeletal motors 
is discussed  in detail in section \ref{sec-intra} while those of matrix 
metalloproteases in the extra-cellular matrix is summarized in 
section \ref{sec-extra}. 
Models of traffic of cells, ants and humans on trails 
are sketched in sections \ref{sec-cellular}, \ref{sec-ants} and \ref{sec-ped}. 
The main conclusions regarding the common trends of modelling the 
traffic-like collective phenomena in diverse systems over a wide range 
of length scales are summarized in section \ref{sec-sum}.

\section{Different types of theoretical approaches}
\label{sec-approach}
                                                                              
First of all, the theoretical approaches can be broadly divided into
two categories: (I) ``Individual-based'' and (II) ``Population-based''. 
The {\it individual-based} models describe the dynamics of the 
individual elements explicitly. Just as ``microscopic'' models of 
matter are formulated in terms of molecular constituents, the 
individual-based models of transport are also developed in terms of 
the constituent elements. Therefore, the individual-based models are 
often referred to as ``microscopic'' models. In contrast, 
in the {\it population-based} models individual elements do not appear 
explicitly and, instead, one considers only the population densities 
(i.e., number of individual elements per unit area or per unit volume).  
The spatio-temporal organization of the elements are emergent collective 
properties that are determined by the responses of the individuals to 
their local environments and the local interactions among the individual 
elements. Therefore, in order to gain a deep understanding of the 
collective phenomena, it is essential to investigate the linkages between 
these two levels of biological organization.

Usually, but not necessarily, space and time are treated as continua
in the population-based models and partial differential equations (PDEs) 
or integro-differential equations are written down for the time-dependent
local collective densities of the elements. The individual-based models 
have been formulated following both continuum and discrete approaches. 
In the continuum formulation of the Lagrangian models, differential 
equations describe the individual trajectories of the elements. 

\subsection{Population-based approaches} 

Suppose $\rho(\vec{x},t)$ is the local density of the population of the 
motile elements at the coarse-grained location $\vec{x}$ at time $t$. 
If the elements are conserved, one can write down an equation of 
continuity for $\rho(\vec{x},t)$:\\ 
\begin{equation}
\frac{\partial \rho}{\partial t} + \vec{\nabla}\cdot \vec{J} = 0.
\label{eq-continuity}
\end{equation}
where $\vec{J}$ is the current density corresponding to the population 
density $\rho$. In addition, depending on the nature of the motile 
elements and their environment, it may be possible to write an analogue 
of the Navier-Stokes equation for the local dynamical variable 
$\vec{v}(\vec{x},t)$.
However, in this review we shall focus almost exclusively on works 
carried out following individual-based approaches.

\subsection{Individual-based approaches} 

For developing an individual-based model, one must first specify the 
{\it state} of each individual element. The dynamical
laws governing the time-evolution of the system must predict the
state of the system at a time $t + \Delta t$, given the corresponding
state at time $t$. The change of state should reflect the response of
the system in terms of movement of the individual elements.

\noindent $\bullet$ Langevin equation: 

One possible framework for the mathematical formulation of such models
is the deterministic Newton's equations for individual elements; each
element is modelled as a ``particle'' subjected to some ``effective
forces'' arising out of its interaction with the other elements.  In
addition, the elements may also experience viscous drag and some
random forces (``noise'') that may be caused by the surrounding
medium.  In that case, instead of the Newton's equation, one can use a
Langevin equation \cite{coffey}. In case the element is an organism
that can think and take decision, capturing inter-element interaction
via effective forces becomes a difficult problem.

For a particle of mass $M$ and instantaneous velocity $v$, the Langevin 
equation describing its motion in one-dimensional space is written as 
\begin{equation}
M \frac{dv}{dt} = F_{\text{ext}} - \Gamma v + F_{br}(t)
\label{eq-lan3}
\end{equation}
where $F_{\text{ext}}$ is the external force acting on the particle, and 
$F_{br}(t)$ is the random force (noise) while the second term on the 
right hand side represents the viscous drag on the particle. In order 
that the average velocity satisfies the Newton equation for a particle 
in a viscous medium, we further assume that 
\begin{equation}
\langle \xi(t)\rangle = 0.
\label{eq-noi1}
\end{equation}
and 
\begin{equation}
\langle\xi(t) \xi(t')\rangle = 2 D T \delta(t-t')
\label{eq-noi2}
\end{equation}
where, $\xi = F_{br}/M$ and, at this level of description, $D$ is a
phenomenological parameter. The prefactor $2$ on the right hand side
of equation (\ref{eq-noi2}) has been chosen for convenience.

An alternative, but equivalent approach is to write down what is now 
generally referred to as a Fokker-Planck equation \cite{risken}. In 
this approach, one deals with a {\it deterministic} partial differential 
equation for a probability density. For example, suppose
$P(\vec{r},\vec{v};t|\vec{r}_0,\vec{v}_0)$
be the conditional probability that, at time $t$, the motile element is
located at $\vec{r}$ and has velocity $\vec{v}$, given that its initial
(i.e., at time $t = 0$) position and velocity were $\vec{r}_0,\vec{v}_0$.
Since the total probability integrated over all space and all velocities
is conserved (i.e,, does not change with time), the probability density
$P$ satisfies an equation of continuity. The probability current density 
gets contribution not only from a diffusive motion of the motile elements 
but also a drift caused by the external force.

Often it turns out that real forces (i.e., forces arising from real 
physical interactions) alone cannot account for the observed dynamics 
of the motile elements; in such situations, ``social forces'' have 
been incorporated in the equation of motion \cite{mogil3,gueron2}. 
However, a priori justification of the forms of such social forces is 
extremely difficult. It is also worth pointing out that, in contrast 
to passive Brownian particles, the motile agents are active Brownian 
particles \cite{schweitzer}.

\noindent $\bullet$ Hybrid approaches: 
                                                                               
Suppose a set of ``particles'', each of which represents a motile 
element, move in a potential field $U[\sigma(x)]$, where the potential 
at any arbitrary location $x$ is determined by the local density 
$\sigma(x)$ of the molecules of a chemical used by the elements for 
communication among themselves. In the case of ants, for example, such 
chemicals are generically called ``pheromone''. Consequently, each 
``particle'' experiences an ``inertial''force $\vec{F}(x) = - \nabla U(x)$. 
Each ``particle'' is also assumed to be subjected to a ``frictional force'' 
where ``friction'' merely parametrizes the tendency of an element to 
continue in a given direction: a smaller ``friction'' implies that the 
element's velocity persists for a longer time in a given direction. The 
motion for the ``particles'' is assumed to be governed by the Langevin 
equation 
\begin{equation}
\ddot{x}  = - \gamma {\dot x} - \nabla U[\sigma(x)] + \eta(t)
\end{equation}
where $\eta(t)$ is a Gaussian white noise with the statistical properties
\begin{equation}
\langle \eta(t)\rangle = 0
\end{equation}
and
\begin{equation}
\langle\eta(t)\eta(t')\rangle = \frac{1}{\beta} \delta(t-t')
\end{equation}
The strength $1/\beta$ of the noise determines the degree of determinacy
with which the particle would follow the gradient of the local potential;
the larger the value of $\beta$ the stronger is the tendency of the
particle to follow the potential gradient.
                                                                               
Thus, the movement of an element may be described as the noisy motion 
of a particle in an ``energy landscape''. However, this energy landscape 
is not static but evolves in response to the motion of the particle as
each particle drops a chemical signal (pheromone) at its own location at 
a rate $g$ per unit time. Assuming that pheromone can diffuse in space 
with a diffusion constant $D$ and evaporate at a rate $\kappa$, the 
equation governing the pheromone field, in one-dimension, is given by
\begin{equation}
\frac{\partial \sigma(x,t)}{\partial t} = D \frac{\partial^{2} 
\sigma(x,t)}{\partial x^{2}} + g \rho(x)- \kappa \sigma(x,t)
\end{equation}
where $\rho(x)$ is the local density of the particles at $x$.
In order to proceed further, one has to assume a specific form of the 
function $U[\sigma(x)]$; one possible form assumed in the case of ants 
\cite{rauch} is 
\begin{equation}
U[\sigma(x)] = - \ln\biggl(1 + \frac{\sigma}{1+\delta \sigma}\biggr)
\end{equation}
where $1/\delta$ is called the capacity.

\noindent $\bullet$ Stochastic cellular automata:

Numerical solution of the Newton-like 
or Langevin-like equations require discretization of both space and time. 
Therefore, the alternative discrete formulations may be used from the 
beginning.
In recent years many individual-based models, however, have been formulated
on discretized space and the temporal evolution of the system in
discrete time steps are prescribed as dynamical update rules using
the language of cellular automata (CA) \cite{wolfram,chopard} or
lattice gas (LG) \cite{marro}. Since each of the individual elements 
may be regarded as an agent, the CA and LG models
are someties also referred to as agent-based models \cite{pnas}.
There are some further advantages in modeling biological systems
with CA and LG. Biologically, it is quite realistic to think in terms of
the way each individual motile element responds to its local environment and
the series of actions they perform. The lack of detailed knowledge of these
behavioral responses is compensated by the rules of CA. Usually, it is much
easier to devise a reasonable set of logic-based rules, instead of cooking
up some effective force for dynamical equations, to capture the behaviour
of the elements. Moreover, because of the high speed of simulations of
CA and LG, a wide range of possibilities can be explored which would be
impossible with more traditional methods based on differential equations.
Most of the models we review in this article are based on CA and LG; 
this modelling strategy focusses mostly on generic features of the system.
The average number of motile elements that arrive at (or depart from) a 
fixed detector site on the track per unit time interval is called the 
{\it flux}. One of the most important transport properties is the relation 
between the flux and the density of the motile elements; a graphical 
representation of this relation is usually referred to as the {\it 
fundamental diagram}. If the motile elements interact mutually only via 
their steric repulsion their average speed $v$ would decrease with 
increasing density because of the hindrance caused by each on the 
following elements. On the other hand, for a given density $c$, the flux 
$J$ is given by $J = c v(c)$, where $v(c)$ is the corresponding average 
speed. At sufficiently low density, the motile elements are well separated 
from each other and, consequently, $v(c)$ is practically independent of 
$c$. Therefore, $J$ is approximately proportional to $c$ if $c$ is very 
small. However, at higher densities the increase of $J$ with $c$ becomes 
slower. At high densitits, the sharp decrease of $v$ with $c$ leads to a 
decrease, rather than increase, of $J$ with increasing $c$. Naturally, the 
fundamental diagram of such a system is expected to exhibit a maxium at 
an intermediate value of the density. 

\section{Asymmetric simple exclusion processes}
\label{sec-asep}

The {\em asymmetric simple exclusion process (ASEP)} \cite{gunterrev}
is a simple particle-hopping model. In the ASEP particles can hop
(with some probability or rate) from one lattice site to a
neighbouring one, but only if the target site is not already occupied
by another particle.  ``Simple Exclusion'' thus refers to the absence
of multiply occupied sites. Generically, it is assumed that the motion
is ``asymmetric'' such that the particles have a preferred direction
of motion.

For a full definition of a model, it is necessary to specify the 
order in which the local rule described above is to be applied to 
the sites. The most common update types are {\em random-sequential 
dynamics} and {\em parallel dynamics}. In the random-sequential case, 
sites are chosen in random order and then updated. In contrast, 
updating for the parallel case is done in a synchronous manner; 
here all the sites are updated at once.

Most often the one-dimensional case is studied, where particles
move along a linear chain of $L$ sites. This is rather natural
for many applications, e.g.\ for modelling highway traffic \cite{css,mahnke}.
If motion is allowed in only one direction (e.g.\ "to the right"), 
the corresponding model is sometimes called Totally Asymmetric 
Simple Exclusion Process (TASEP). The probability of motion from 
site $j$ to site $j+1$ will be denoted by $p$; in the simplest case, 
where all the sites are treated on equal footing, $p$ is assumed 
to be independent of the position of the particle.

\begin{figure}[tb]
\begin{center}
\includegraphics[angle=-90,width=0.5\columnwidth]{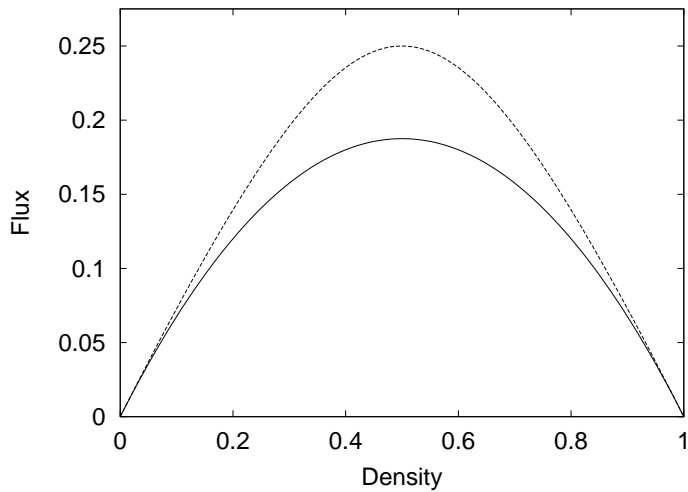}
\end{center}
\caption{Fundamental diagrams of the TASEP with periodic boundary
conditions and (a) random-sequential updating (solid curve)
(b) parallel updating (dashed curve), both for $p = 0.75$.}
\label{fig-nsfunda}
\end{figure}

For such driven diffusive systems the boundary conditions turn out
to be crucial. If periodic boundary conditions are imposed, i.e., 
the sites $1$ and $L$ are made nearest-neibours of each other, all 
the sites are treated on the same footing. For this system the 
fundamental diagram has been derived exactly both in the cases of 
parallel and random-sequential updating rules \cite{css}; these 
are shown graphically in fig.\ref{fig-nsfunda}. 

If the boundaries are open, then a particle can enter from a reservoir 
and occupy the leftmost site ($j=1$), with probability $\alpha$, if 
this site is empty. In this system a particle that occupies the 
rightmost site ($j=L$) can exit with probability $\beta$.

\begin{figure}[ht]
\centerline{\epsfig{figure=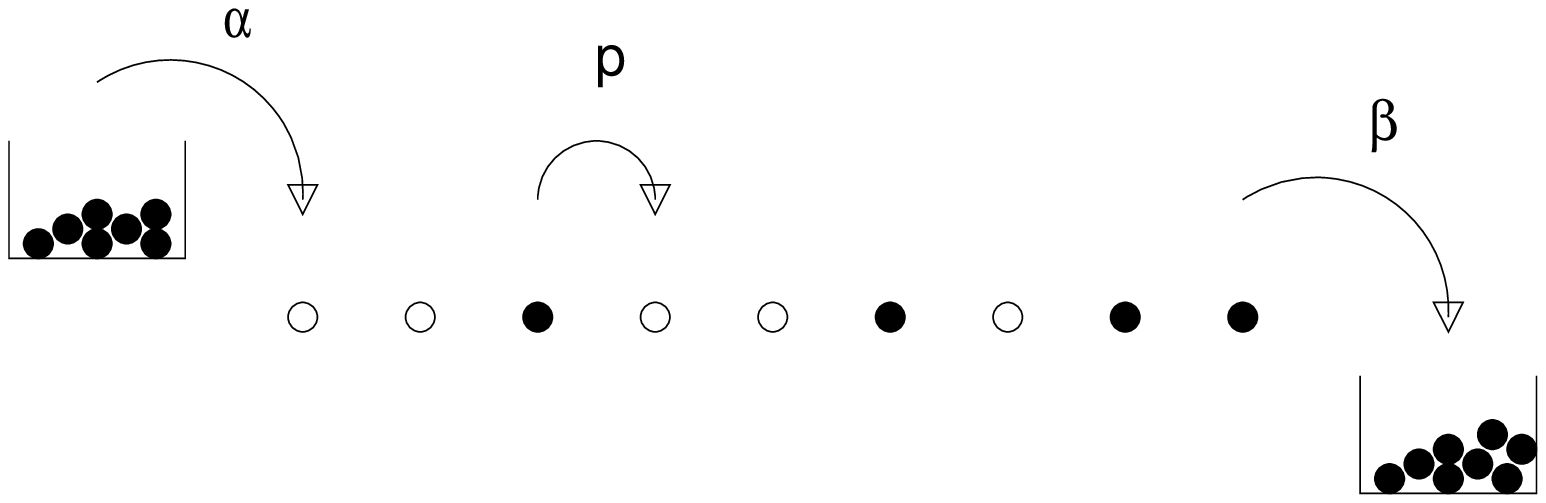,width=7cm}\qquad
\epsfig{figure=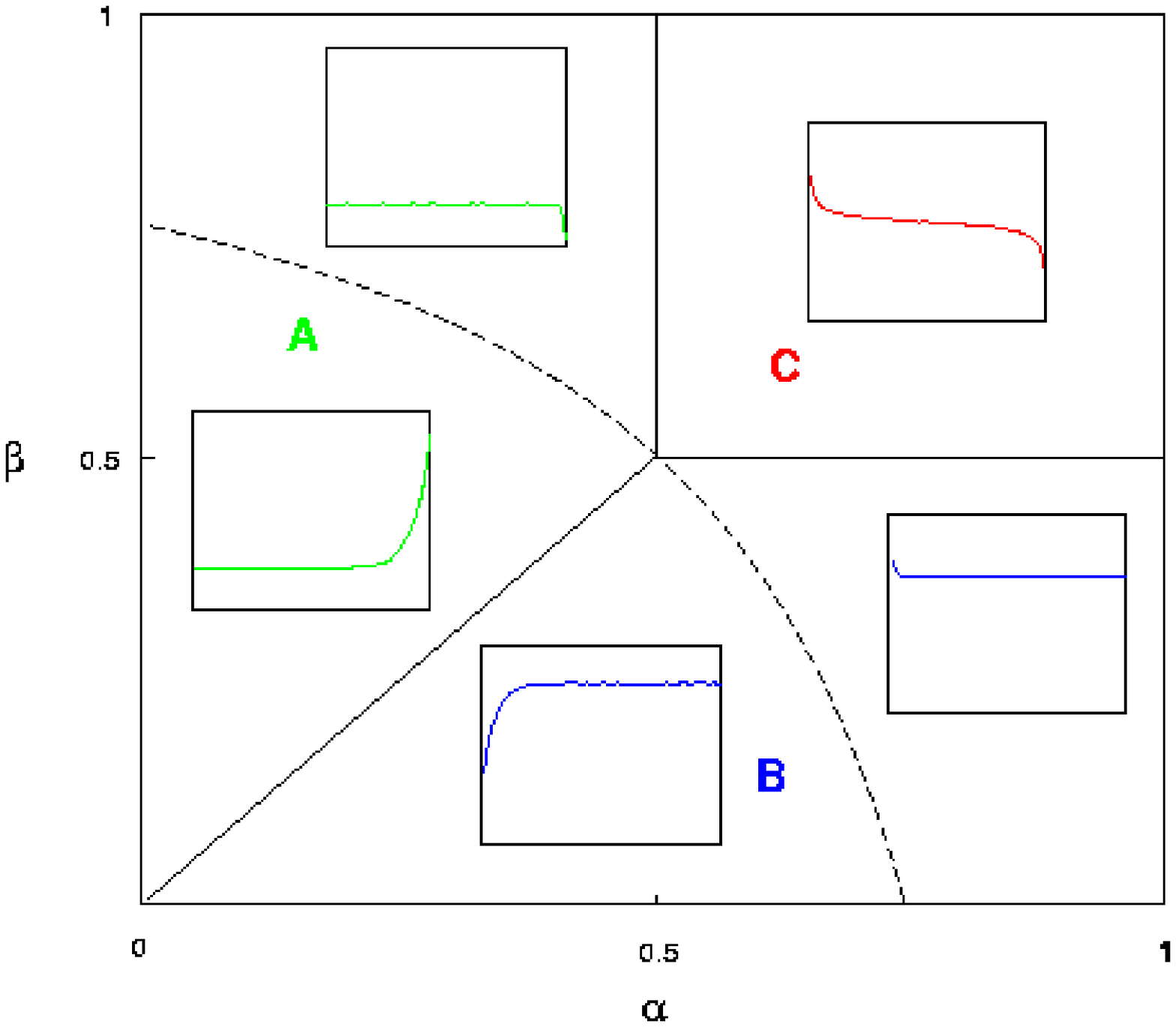,width=8cm}}
\caption{\protect{Processes allowed in the ASEP (left);
Phase diagram of the ASEP for parallel dynamics (right) and $p = 0.75$. 
The insets show the typical shape of the density profiles in the 
corresponding region. The broken line indicates points with
a flat density profile.}}
\label{fig_asepphase}
\end{figure}

The ASEP has been studied extensively in recent years and is now well
understood (see e.g.\ \cite{evansAlten,gunterrev} and references therein). 
In fact its stationary state for different dynamics
can be obtained exactly \cite{derrida93,schdom,rsss,ERS,degier}. 
It shows an interesting phase diagram
(see Fig.~\ref{fig_asepphase}) and is the prototype for so-called 
boundary-induced phase transitions \cite{krug}.

Fig.~\ref{fig_asepphase} shows the generic form of the phase diagram
obtained by varying the boundary rates $\alpha$ and $\beta$. 
One can distinguish three phases,
namely (A) a low-density phase  ($\alpha < \beta,\alpha_c(p)$), 
(B) a high-density phase ($\beta < \alpha,\beta_c(p)$) and
(C) a maximal-current phase ($\alpha>\alpha_c(p)$ and $\beta>\beta_c(p)$). 
The appearance of these three phases
can easily be understood. In the low-density phase the current
depends only on the input rate $\alpha$. The input is less efficient
than the transport in the bulk of the system or the output and
therefore dominates the behaviour of the whole system. In the
high-density phase the output is the least efficient part of
the system. Therefore the current depends only on $\beta$. In
the maximal current phase, input and output are more efficient
than the transport in the bulk of the system. Here the current
has reached the largest possible value corresponding to the maximum
of the fundamental diagram of the periodic system.

%

Mean-field theory \cite{macdonald}
predicts the existence of a shock or domain wall that separates a
macroscopic low-density region at the start-end of the chain from a
macroscopic high-density region at the stop-end. The exact solution
\cite{derrida93,schdom}, on the other hand, gives a linear increasing
density profile. These two results do not contradict each other
since the sharp domain wall, due to current fluctuations, performs
a random walk along the lattice. The mean-field result therefore
corresponds to a snapshot at a given time whereas the exact solution
averages over all possible positions of the shock.

In \cite{Kolo98} a nice physical picture has been developed which
explains the structure of the phase diagram not only qualitatively,
but also (at least partially) quantitatively. It remains 
correct even for more sophisticated models \cite{popkov}.
It relates the phase boundaries to properties of the periodic system
which can be derived from the fundamental diagram, namely the
so-called shock velocity $v_s=\frac{J(\rho^+)-J(\rho^-)}{\rho^+ - \rho^-}$ 
and the collective velocity $v_c=\frac{\partial J(\rho)}{\partial \rho}$.
$v_{s}$ is the velocity of a 'domain-wall' which in nonequilibrium 
systems denotes an object connecting two possible stationary states. 
Here these stationary states have densities $\rho^+ $ and $\rho^-$,
respectively. The collective velocity $v_c$ describes the velocity
of the center-of-mass of a local perturbation in a homogeneous,
stationary background of density $\rho$.
The phase diagram of the open system is then completely determined by
the fundamental diagram of the periodic system through an
extremal-current principle \cite{popkov2} and therefore independent
of the microscopic dynamics of the model.

\begin{figure}[ht]
\begin{center}
\includegraphics[width=0.5\columnwidth]{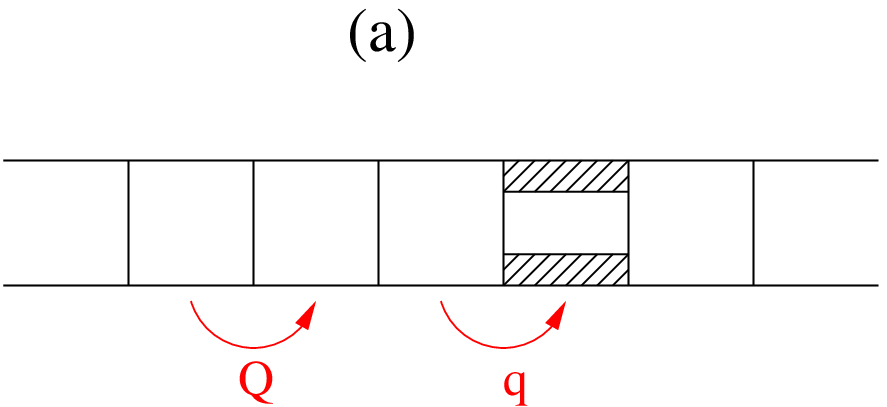}
\includegraphics[width=0.5\columnwidth]{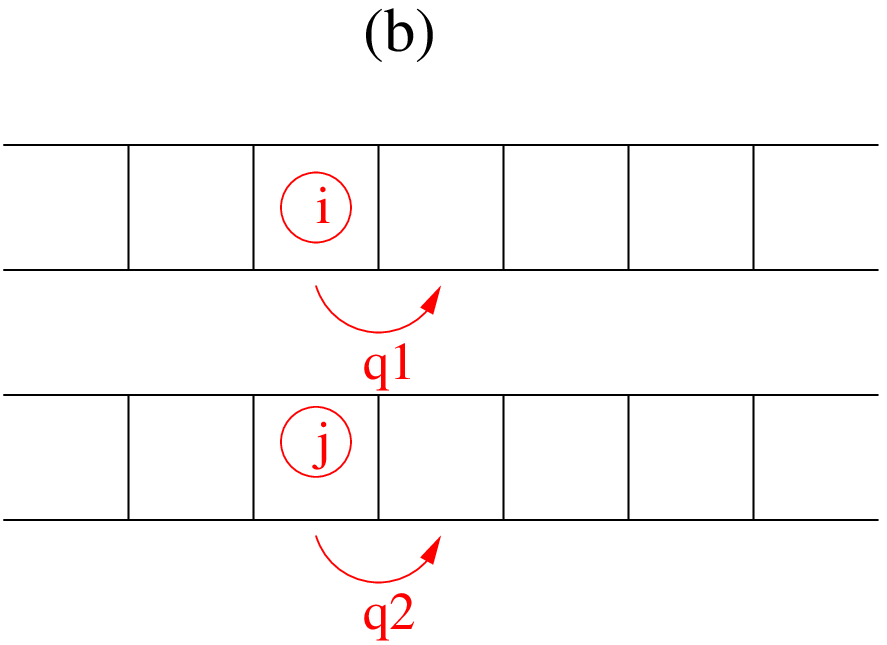}
\includegraphics[width=0.5\columnwidth]{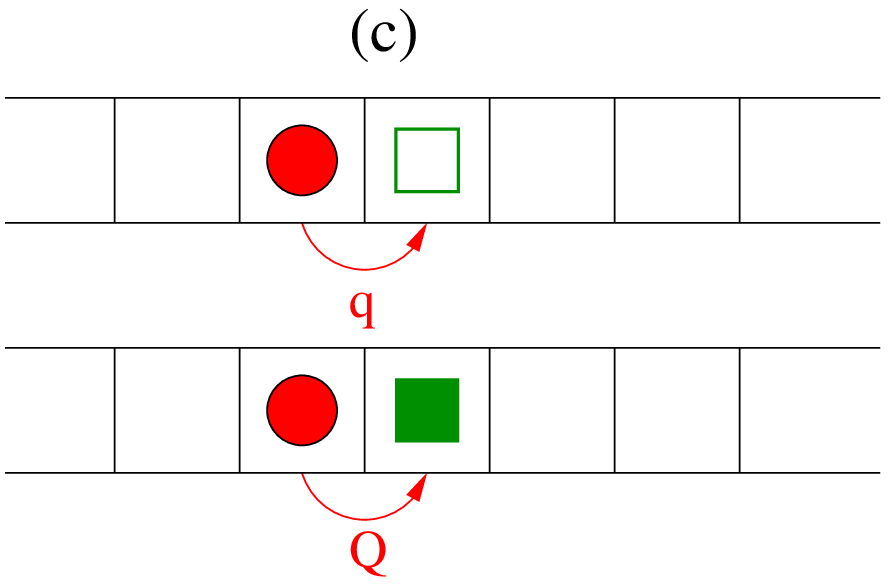}
\end{center}
\caption{Schematic representation of the different types of randomness
in particle-hopping models. In (a) the randomness is associated with
the track; the hopping probability $q$ at the bottleneck (partially
hatched region) is smaller than the normal hopping probability $Q$,
In (b) the randomness is associated with the particles; $q_1$ and $q_2$
being the time-independent hopping probabilities of the particles $i$
and $j$, respectively. In (c) the randomness arises from the coupling
of the dynamics of the hopping particles (filled circle) with another
species of particles that represent specific type of signal molecules; 
the two possible states of the latter are represented by open and filled 
squares.}
\label{fig-disorder}
\end{figure}
                                                                                
\subsection{Modelling randomness in ASEP-type models}

At least three different types of randomness of the hopping rates have 
been considered so far in the context of the ASEP-type models 
\cite{janowsky,tripathy,goldspeer,krugdis,evansdis,ktitarev,igloi}. 

(a) First, the randomness may be associated with the {\it track} on which 
the motile elements move \cite{janowsky,tripathy,goldspeer,igloi}; typical 
examples are the bottlenecks created, in intra-cellular transport in 
neurons, by {\it tau}, a microtubule-associated protein 
\cite{mandeltau1,mandeltau2}. The inhomogeneities of some DNA and m-RNA 
strands can be well approximated as random \cite{kafri} and, hence, the 
randomess of hopping of the motile elements on the nucleotide-based tracks.
As shown schematically in fig.\ref{fig-disorder}(a), normal hopping 
probability at unblocked sites is $Q$ whereas that at the bottleneck is 
$q$ ($q < Q$). This type of randomness in the hopping probabilities,  
which may be treated as quenched (i.e., time-independent or ``frozen'') 
defect of the track, leads to interesting phase-segregation phenomena 
(see \cite{css} for a review).
                                                                                
(b) The second type of randomness is associated with the hopping {\it
motile elements} \cite{krugdis,evansdis,ktitarev,igloi}, rather than 
with the track. For example, the normal hopping probabilities of the  
motile elements may vary randomly from one element to another 
(see fig.\ref{fig-disorder}(b)), e.g., in randomly mutated kinesins 
\cite{traffic,hirotaked}; the hopping rate of each motile element is, 
however, ``quenched'', i.e., independent of time. In this case, the 
system is known to be exhibit coarsening of queues of the motile elements 
and the phenomenon has some formal similarities with Bose-Einstein 
condensation (reviewed in \cite{css}).
Note that in case of the randomness of type (a), the hopping 
probability depends only on the spatial location on the track, independent 
of the identity of the hopping motile element. On the other hand, in the 
case of randomness of type (b), the hopping probability depends on the 
hopping motile element, irrespective of its spatial location on the track. 

(c) In contrast to the two types of randomness ((a) and (b)) considered 
above, the randomness in the hopping probabilities of the motile elements 
in some situations arises from the coupling of their dynamics with that 
of another non-conserved dynamical variable. For example, the hopping 
probability of a motile element may depend on the presence or absence of 
a specific type of signal molecule in front of it 
(see fig.\ref{fig-disorder}(c)); such situations arise in traffic of ants 
whose movements are strongly dependent on the presence or absence of 
pheromone on the trail ahead of them. Therefore, in such models with 
periodic boundary conditions, a given motile element may hop from the 
same site, at different times, with different hopping probabilities.

\section{Generic mechanisms of single molecular motor}
\label{sec-single}

Two extremely idealized mechanisms of motility of single-motors have 
been developed in the literature. The {\it power-stroke} mechanism 
is analogous to the power strokes that drive macrscopic motors. On 
the other hand, the {\it Brownian ratchet} mechanism is unique to 
the microscopic molecular motors.

\begin{figure}[ht]
\begin{center}
\includegraphics[angle=-90,width=0.65\columnwidth]{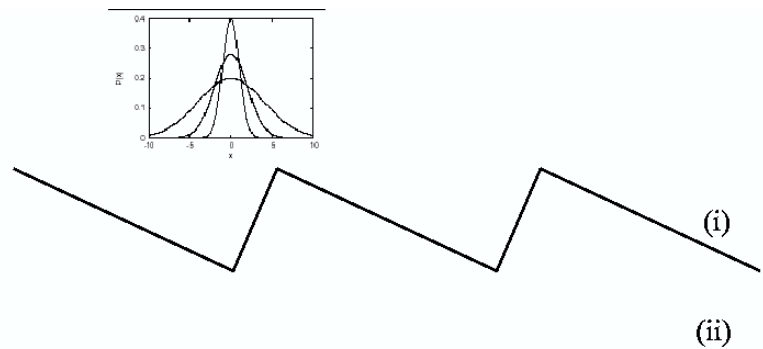}
\end{center}
\vspace{-0.9cm}
\caption{The two forms of the time-dependent potential used for
implementing the Brownian ratchet mechanism.
(Copyright: Indrani Chowdhury; reproduced with permission).
}
\label{fig-ratgauss}
\end{figure}

\begin{figure}[ht]
\begin{center}
\includegraphics[angle=-90,width=0.5\columnwidth]{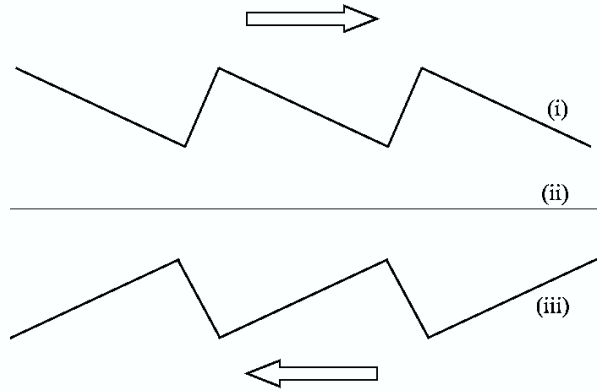}
\end{center}
\caption{The direction of the motion of the particle in a Brownian ratchet,  
is determined by the form of the spatial asymmetry of the potential in each 
period when observed over times much longer than the time period of switching 
between the two forms of the potential.
(Copyright: Indrani Chowdhury; reproduced with permission).
}
\label{fig-ratbidir}
\end{figure}

Let us now consider a Brownian particle subjected to a {\it time-dependent}
potential, in addition to the viscous drag (or, frictional force). The
potential switches between the two forms (i) and (ii) shown in
fig.\ref{fig-ratgauss}. The sawtooth form (i) is spatially {\it periodic}
where each period has an {\it asymmetric} shape. In contrast,
the form (ii) is flat so that the particle does not experience any
external force imposed on it when the potential has the form (ii).
Note that, in the left part of each well in (i) the particle
experiences a rightward force whereas in the right part of the same
well it is subjected to a leftward force. Moreover, the spatially
averaged force experienced by the particle in each well of length
$\ell$ is
\begin{eqnarray}
\langle F\rangle  = - \frac{1}{\ell}\int_0^{\ell} \biggl(
\frac{\partial U}{\partial x}\biggr)dx 
= U(0) - U(\ell) = 0
\end{eqnarray}
because of the spatially periodic form of the potential (i). What
makes this problem so interesting is that, in spite of vanishing average
force acting on it, the particle can still exhibit directed, albeit
noisy, rightward motion.
                                                                               
In order to understand the underlying physical principles, let us
assume that initially the potential has the shape (i) and the
particle is located at a point on the line that corresponds to the
bottom of a well. Now the potential is switched off so that it makes
a transition to the form (ii). Immediately, the free particle begins
to execute a Brownian motion and the corresponding Gaussian profile
of the probability distribution begins to spread with the passage of
time.  If the potential is again switched on before the Gaussian profile
gets enough time for spreading beyond the original well, the particle
will return to its original initial position. But, if the period
during which the potential remains off is sufficiently long, so that
the Gaussian probability distribution has a non-vanishing tail
overlapping with the neighbouring well on the right side of the
original well, then there is a small non-vanishing probability that
the particle will move forward towards right by one period when the
potential is switched on. In the case of cytoskeleton-based motors 
like kinesin and dynein, this energy is supplied by the hydrolysis 
of ATP molecules to ADP; thus, the mechanical movement is coupled to 
a chemical reaction.
                                                                               
In this mechanism, the particle moves forward not because of any
force imposed on it but because of its Brownian motion. The system
is, however, not in equilibrium because energy is pumped into it
during every period in switching the potential between the two
forms. In other words, the system works as a rectifier where the
Brownian motion, in principle, could have given rise to both
forward and backward movements of the particle in the multiples of
$\ell$, but the backward motion of the particle is suppressed by
a combination of (a) the time dependence and (b) spatial asymmetry
(in form (i)) of the potential. In fact, the direction of motion
of the particle can be reversed by replacing the potential (i) by
the potential (iii) shown in fig.\ref{fig-ratbidir}. The spatial 
asymmetry of the sawtooth potential arises from the polar nature of 
the microtubule and actin filamentary tracks.

The mechanism of directional movement discussed above is called a
Brownian ratchet \cite{julicher,reimann}. 
The concept of Brownian ratchet was popularized by Feynman through his
lectures \cite{feynman} although, historically, it was introduced by
Smoluchowski \cite{smolurat}.

\subsection{Modelling defects and disorder in Brownian ratchets} 

Effects of quenched (i.e., time-independent) disorder on the properties 
of Brownian ratchets have been considered by several authors 
\cite{harms,marchesoni,family,kafri}.
Quenched disorder can arise in Brownian ratchets, for example, from \\ 
(i) random variation of the heights (or depths) of the sawtooth potential 
from one site to another where all the sawteeth have the same type of 
asymmetry,\\
(i) a random mixture of forward and reversed sawteeth where the heights 
of all the sawteeth is identical. 
The nature of disorder in real molecular motors, even if driven by a 
Brownian-ratchet mechanism, may be a combination of these two types of 
idealized disorder.

Suppose $\omega$ is the frequency of both the transitions from (i) to 
(ii) and (ii) to (i) forms of the potential. Also, let $P_D$ be the 
probability of finding a defect, i.e., a reversed sawtooth in case (ii). 
In that case, the effective drift and effective diffusion coefficient 
exhibit three different regions on the $P_D-\omega$ phase diagram 
\cite{harms} including some anomalous behaviour.

\section{Intracellular transport: nucleotide-based motors} 
\label{sub-protein}

Helicases and polymerases are the two classes of nucleotide-based 
motors that have been the main focus of experimental investigations. 
In this section, we discuss only the motion of the ribosome along 
the m-RNA track.  Historically, this problem is one of the first 
where TASP-like model was successfully applied to a biological system.

The synthesis of proteins and polypetids in a living cell is a complex
process. Special machines, so-called {\em ribosomes}, translate the 
genetic information `stored' in the {\em messenger-RNA (mRNA)} into a 
program for the synthesis of a protein. mRNA is a long (linear) molecule 
made up of a sequence of triplets of nucleotides; each triplet is called 
a {\it codon}. The genetic information is encoded in the sequence of 
codons. A ribosome, that first gets attached to the mRNA chain, ``reads'' 
the codons as it moves along the mRNA chain, recruits the corresponding 
amino acids and assembles these amino acids in the sequence so as to 
synthesize the protein for which the ``construction plan'' was stored in 
the mRNA. After executing the synthesis as per the plan, it gets 
detached from the mRNA. Thus, the process of ``translation'' of genetic 
information stored in mRNA consists of three steps: (i) {\em initiation}: 
attachment of a ribosome at the ``start'' end of the mRNA, (ii) 
{\em elongation}: of the polypeptide (protein) as the ribosome moves along 
the mRNA, and (iii) {\em termination}: ribosome gets detached from the 
mRNA when it reaches the ``stop'' codon.

Let us denote each of the successive codons by the successive sites of 
a one-dimensional lattice where the first and the last sites correspond 
to the start and stop codons. The ribosomes are much bigger (20-30 times) 
than the codons. Therefore, neighbouring ribosomes attached to the same 
mRNA can not read the same information or overtake each other. In other 
words, any given site on the lattice may be covered by a single ribosome 
or none. Let us represent each ribosome by a rigid rod of length $L_{r}$. 
If the rod representing the  ribosome has its left edge  attached to 
the i-th site of the lattice, it is allowed to move to the right by one 
lattice spacing, i.e., its left edge moves to the site $i+1$ provided the 
site $i+L_r$ is empty. In the special case $L_r = 1$ this model reduced 
to the TASEP. Although the model was originally proposed in the late 
sixties \cite{macdonald}, significant progress in its analytical 
treatment for the general case of arbitrary $L_r$ could be made only 
three decades later; even the effects of quenched disorder has also been 
considered in the recent literature 
\cite{shaw1,shaw2,shaw3,lakatos1,lakatos2}.

\begin{figure}[ht]
\centerline{\psfig{figure=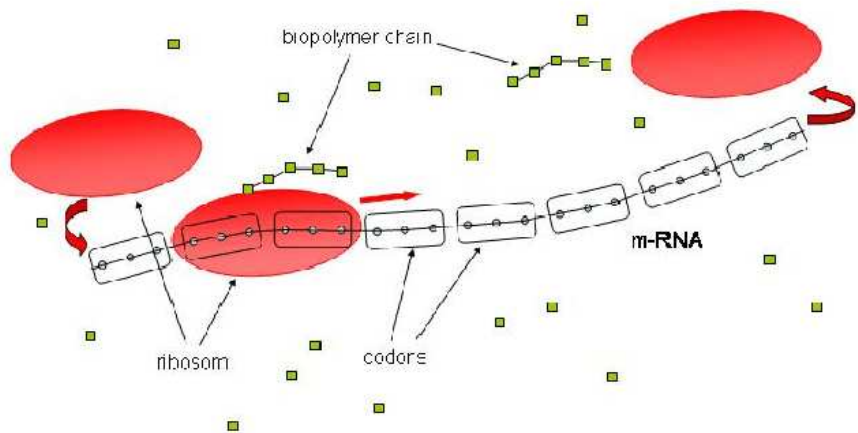,width=12cm}}
\caption{\protect{The process of biopolymerization: Ribosomes attach to 
mRNA and read the construction plan for a biopolymer which is stored in 
the genetic code formed by the sequence of codons. None of the codons 
can be read by more than one ribosome simultaneously.
}}
\label{fig_mRNA}
\end{figure}

As mentioned above, a ribosom is much bigger
than a base triplet. However, modifying the ASEP by taking into
account particles that occupy more than one lattice site does not
change the structure of phase diagram \cite{macdonald}. Physically
this can be understood from the domain-wall picture and the
extremal-current principle \cite{gunterrev,schdom}.

\section{Intracellular transport: cytoskeleton-based motors} 
\label{sec-intra}

Intracellular transport is carried by molecular motors which are proteins 
that can directly convert the chemical energy into mechanical energy 
required for their movement along filaments constituting what is known 
as the cytoskeleton \cite{howard,schliwa}. Three superfamilities of 
these motors are kinesin, dynein and myosin. Members of the majority of 
the familities have two heads whereas only a few families have 
single-headed members. Most of the kinesins and dyneins are like porters 
in the sense that these move over long distances along the filamentary 
tracks without getting completely detached; such motors are called 
{\it processive}. On the other hand, the conventional myosins and a few  
unconventional ones are nonprocessive; they are like rowers. But, a 
few families of unconventional myosins are processive. 

These cytoskeleton-based molecular motors play crucially important 
biological functions in axonal transport in neurons, intra-flagellar 
transport in eukaryotic flagella, etc. The relation between the 
architectural design of these motors and their transport function has 
been investigated both experimentally and theoretically for quite some 
time \cite{osterrev,fisher,astu1}. 

However, in this review we shall focus mostly on the effects of mutual 
interactions (competition as well as cooperation) of these motors on 
their collective spatio-temporal organisation and the biomedical 
implications of such organisations. Often a single microtubule (MT) is 
used simultaneously by many motors and, in such circumstances, the 
inter-motor interactions cannot be ignored. Fundamental understanding 
of these collective physical phenomena may also expose the causes of 
motor-related diseases (e.g., Alzheimer's disease) 
\cite{traffic,hirotaked,mandeld,goldstein} thereby helping, possibly, 
also in their control and cure. The bio-molecular motors have opened up 
a new frontier of applied research- ``bio-nanotechnology''. A clear 
understanding of the mechanisms of these natural machines will give us 
clue as to the possible design principles that can be utilized to 
synthesize artificial nanomachines.

Derenyi and collaborators \cite{derenyi1,derenyi2} developed 
one-dimensional models of interacting Brownian motors, each of which 
is subjected to a time-dependent potential of the form shown in 
fig.\ref{fig-ratgauss}. They modelled each motor as a {\it rigid rod} 
and formulated the dynamics through Langevin equations of the form 
(\ref{eq-lan3}) for each such rod assuming the validity of the 
overdamped limit; the mutual interactions of the rods were incorporated 
through the mutual exclusion. However, in this section we shall focus 
attention on those models where the dynamics is formulated in terms of 
``rules'' for undating in discrete time steps.

\subsection{TASEP-like generic models of molecular motor traffic} 

The model considered by Aghababaie et al.\cite{menon1} is not based on 
TASEP, but its dynamics is a combination of Brownian ratchet and update 
rules in discrete time steps. More precisely, this model is a 
generalization of TASEP, rather than TASEP, where the hopping 
probabilities are obtained from the local potential which itself is 
time-dependent and is assumed to have the form shown in 
fig.\ref{fig-ratgauss}. 

In this model, the filamentary track is discretized in the spirit of 
the particle-hopping models described above and the motors are 
represented by {\it field-driven} particles; no site can accomodate 
more than one particle at a time. Each time step consists of either 
an attempt of a particle to hop to a neighbouring site or an attempt 
that can result in switching of the potential from flat to sawtooth 
form or vice-versa. Both forward and backward movement of the particles 
are possible and the hopping probability of every particle is computed 
from the instantaneous local potential. However, neither attachment of 
new particles nor complete detachment of existing particles were allowed.

The fundamental diagram of the model \cite{menon1}, computed imposing 
periodic boundary conditions, is very similar to those of TASEP. This 
observation indicates that further simplification of the model proposed 
in ref.\cite{menon1} is possible to develope a minimal model for interacting
molecular motors. Indeed, the detailed Brownian ratchet mechanism, which 
leads to a noisy forward-directed movement of the {\it field-driven} 
particles in the model of Aghababaie et al. \cite{menon1}, is replaced 
in some of the more recent theoretical models 
\cite{lipo1,lipo2,lipo3,lipo4,lipo5,lipo6,lipo7,frey,santen1,santen2,popkov1} 
by a TASEP-like probabilitic forward hopping of {\it self-driven} 
particles. In these simplied versions, none of the particles is allowed 
to hop backward and the forward hopping probability is assumed to capture 
most of the effects of biochemical cycle of the enzymatic activity of the 
motor. The explicit dynamics of the model is essentially an extension 
of that of the asymmetric simple exclusion processes (ASEP) 
\cite{sz,gunterrev} (see section~\ref{sec-asep}) that includes, in addition, 
Langmuir-like kinetics of adsorption and desorption of the motors.

\noindent $\bullet$ Model proposed by Parmeggiani et al.

\begin{figure}[ht]
\begin{center}
\includegraphics[angle=-90,width=0.4\textwidth]{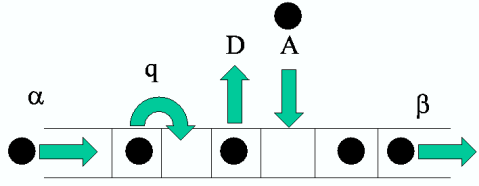}\\
\end{center}
\caption{A schematic description of the TASEP-like model introduced in 
ref.\cite{frey} for molecular motor traffic. Just as in TASEP, the 
motors are allowed to hop forward, with probability $q$. In addition, 
the motors can also get ``attached'' to an empty lattice site, with 
probability $A$, and ``detached'' from an occupied site, with probability 
$D$ from any site except the end points; the rate of attachment at 
the entry point on the left is $\alpha$ while that at the exit point 
on the right is $\beta$.
}
\label{fig-frey}
\end{figure}

In the model of Parmeggiani et al. \cite{frey}, the molecular motors 
are represented by particles whereas the sites for the binding of the
motors with the cytoskeletal tracks (e.g., microtubules) are represented
by a one-dimensional discrete lattice.  Just as in TASEP, the motors are
allowed to hop forward, with probability $q$, provided the site in
front is empty. However, unlike TASEP, the particles can also get
``attached'' to an empty lattice site, with probability $A$,
and ``detached'' from an occupied site, with probability $D$
(see fig.\ref{fig-frey}) from any site except the end points. The state
of the system was updated in a random-sequential manner.

Carrying out Monte-Carlo simulations of the model, applying open
boundary conditions, Parmeggiani et al.\cite{frey} demonstrated a
novel phase where low and high density regimes, separated from each
other by domain walls, coexist \cite{santen1,santen2}. Using a 
mean-field theory (MFT), they interpreted this spatial organization 
as traffic jam of molecular motors. This model has interesting 
mathematical properties \cite{sutapa} which are of fundamental interest 
in statistical physics but are beyond the scope of this review.

\noindent $\bullet$ Model proposed by Klumpp et al. 

\begin{figure}[ht]
\begin{center}
\includegraphics[angle=-90,width=0.4\textwidth]{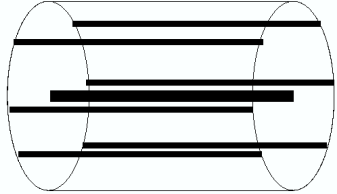}\\
\end{center}
\caption{A schematic description of the TASEP-like model introduced in 
ref.\cite{lipo1} for molecular motor traffic in a tubular geometry 
that mimics axonal transport. The thick axial line represents the 
microtubule track whereas the thin lines surrounding it merely indicate 
the non-directed diffusive channels of transport along the tube.
}
\label{fig-lipowsky}
\end{figure}

\begin{figure}[ht]
\begin{center}
\includegraphics[angle=-90,width=0.4\textwidth]{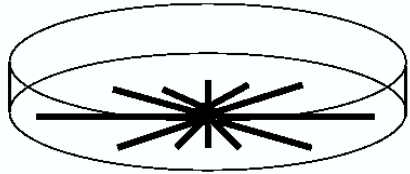}\\
\end{center}
\caption{A schematic description of the TASEP-like model introduced in 
ref.\cite{lipo7} for intra-cellular molecular motor traffic in the 
radial geometry that mimics an aster. 
}
\label{fig-lipo2}
\end{figure}

A cylindrical geometry of the model system (see fig.\ref{fig-lipowsky}) 
was considered by Lipowsky, Klumpp and collaborators 
\cite{lipo1,lipo2,lipo3,lipo4,lipo5,lipo6,lipo7} to mimic the 
microtubule tracks in typical tubular neurons. The microtubule filament 
was assumed to form the axis of the cylinder whereas the free space 
surrounding the axis was assumed to consist of $N_{ch}$ channels each 
of which was discretized in the spirit of lattice gas models. They 
studied concentration profiles and the current of free motors as well 
as those bound to the filament by imposing a few different types of 
boundary conditions. This model enables one to incorporate the effects 
of exchange of populations between two groups, namely, motors bound to 
the axial filament and motors which move diffusively in the cylinder. 
They have also compared the results of these investigations with the 
corresponding results obtained in a different geometry where the filaments 
spread out radially from a central point (see fig.\ref{fig-lipo2}).

\noindent $\bullet$ Model proposed by Klein et al. 

It is well known that, in addition to generating forces and carrying 
cargoes, cytoskeletal motors can also depolymerize the filamentary 
track on which they move processively. A model for such filament 
depolymerization process has been developed by Klein et al.\cite{klein} 
by extending the model of intra-cellular traffic proposed earlier by 
Parmeggiani et al. \cite{frey}. 

\begin{figure}[ht]
\begin{center}
\includegraphics[angle=-90,width=0.4\textwidth]{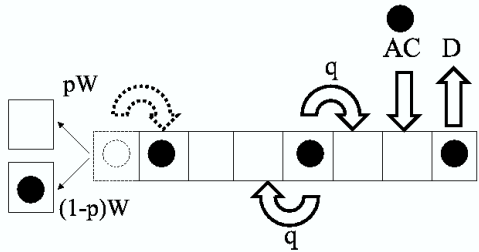}\\
\end{center}
\caption{A schematic description of the model suggested by Klein et al. 
\cite{klein} for motor induced depolymerization of cytoskeletal 
filaments. The lattice site at the tip of a filament is removed with a 
probability $W$ per unit time provided it is occupied by a motor;
the motor remains attached to the newly exposed tip of the filament
with probability $p$ (or remains bound with the removed site with
probability $1-p$). The other symbols and procsses are identical to 
those shown in fig.\ref{fig-frey}.
}
\label{fig-klein}
\end{figure}

The model of Klein et al.\cite{klein} is shown schematically in fig.
\ref{fig-klein}. The novel feature of this model, in contrast to the 
similar models \cite{frey,lipo7,santen1} of intracellular 
transport, is that the lattice site at the tip of a filament is removed 
with a probability $W$ per unit time provided it is occupied by a motor; 
the motor remains attached to the newly exposed tip of the filament 
with probability $p$ (or remains bound with the removed site with 
probability $1-p$). Thus, $p$ may be taken as a measure of the 
processivity of the motors. This model clearly demonstrated a dynamic 
accumulation of the motors at the tip of the filament arising from the 
processivity; a motor which was bound to the depolymerizing monomer at 
the tip of the filament is captured by the monomer at the newly exposed 
tip.

\noindent $\bullet$ Model proposed by Kruse and Sekimoto:

\begin{figure}[t]
\begin{center}
\vspace{0.5cm}
\includegraphics[angle=-90,width=0.7\columnwidth]{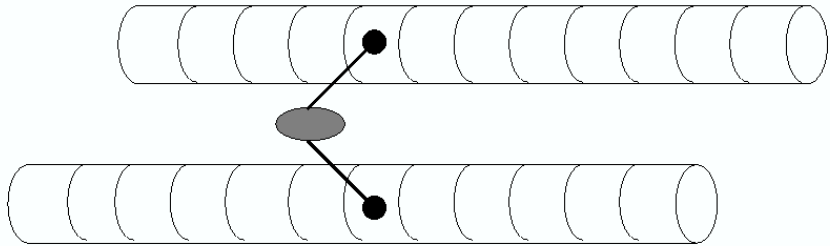}
\end{center}
\caption{Schematic representation of Kruse and Sekimito's model 
\cite{sekimoto} for the motor-induced sliding of two cytoskeletal 
filaments. The cylinders represent the two filamentary tracks. 
The dark circles represent the two heads of a motor 
which are capable of binding on two binding sites on the two 
filaments provided the two binding sites are closest neighbours 
of each other. Each head can move forward following a TASEP-like 
rule and every movement of this type causes sliding of the two 
filaments by one single unit.
}
\label{fig-sekimoto}
\end{figure}

Kruse and Sekimoto \cite{sekimoto} proposed a particle-hopping model 
for motor-induced relative sliding of two filamentary motor tracks. 
The model is shown schematically in fig.\ref{fig-sekimoto}. Each 
of the two-headed motors is assumed to consist of two particles 
connected to a common neck and are capable of binding with two 
filaments provided the two binding sites are closest neighbours as 
shown in the figure. Each particle can move forward following a 
TASEP-like rule and every movement of this type causes sliding of the 
two filaments by one single unit. The most important result of this 
investigation is that the average relative velocity of the filaments 
is a non-monotonic function of the concentration of the motors.

\subsection{Traffic of interacting single-headed motors KIF1A} 

The models of intracellular traffic described so far are essentially 
extensions of the asymmetric simple exclusion processes (ASEP) 
\cite{sz,gunterrev} that includes Langmuir-like kinetics of adsorption 
and desorption of the motors. In reality, a motor protein is an enzyme 
whose mechanical movement is loosely coupled with its biochemical cycle. 
In a recent work \cite{nishietal}, we have considered specifically the
{\it single-headed} kinesin motor, KIF1A 
\cite{okada1,okada2,okada3,Nitta,unpub};
the movement of a single KIF1A motor was modelled earlier with a Brownian 
ratchet mechanism \cite{julicher,reimann}. In contrast to the earlier 
models \cite{frey,santen1,popkov1,lipo7} of molecular motor traffic,
which take into account only the mutual interactions of the motors, our
model explicitly incorporates also the Brownian ratchet mechanism of
individual KIF1A motors, including its biochemical cycle that involves
{\it adenosine triphosphate(ATP) hydrolysis}.

The ASEP-like models successfully explain the occurrence of shocks.
But since most of the bio-chemistry is captured in these models through
a single effective hopping rate, it is difficult to make direct
quantitative comparison with experimental data which depend on such
chemical processes. In contrast, the model we proposed in ref.
\cite{nishietal} incorporates the essential steps in the biochemical 
processes of KIF1A as well as their mutual interactions and involves 
parameters that have one-to-one correspondence with experimentally 
controllable quantities.

\begin{figure}[ht]
\begin{center}
\includegraphics[width=0.4\textwidth]{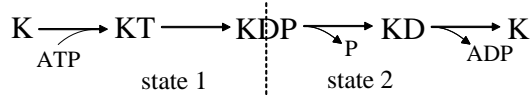}\\
\end{center}
\caption{The biochemical and mechanical states of a single KIF1A motor.
In the chemical states on the left of the dotted line, KIF1A binds to
a fixed position on the MT (state 1), while in those on the right KIF1A
diffuses along the MT track (state 2). At the transition from state 1 to
2, KIF1A detaches from the MT.}
\label{fig-cycle}
\end{figure}

The biochemical processes of kinesin-type molecular motors can be
described by the four states model shown in Fig.~\ref{fig-cycle}
\cite{okada1,Nitta}: bare kinesin (K), kinesin bound with ATP (KT),
kinesin bound with the products of hydrolysis, i.e., adenosine
diphosphate(ADP) and phosphate (KDP), and, finally, kinesin bound with
ADP (KD) after releasing phosphate. Recent experiments \cite{okada1,Nitta}
revealed that both K and KT bind to the MT in a stereotypic manner
(historically called ``strongly bound state'', and here we refer to
this mechanical state as ``state 1''). KDP has a very short lifetime and
the release of phosphate transiently detaches kinesin from MT \cite{Nitta}.
Then, KD re-binds to the MT and executes Brownian motion along the
track (historically called ``weakly bound state'', and here referred to
as ``state 2''). Finally, KD releases ADP when it steps forward to the
next binding site on the MT utilizing a Brownian ratchet mechanism, and
thereby returns to the state K.

\begin{figure}[ht]
\begin{center}
\includegraphics[width=0.45\textwidth]{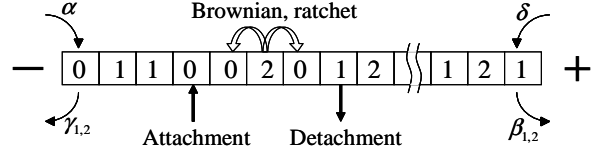}\\
\end{center}
\caption{A 3-state model for molecular motors moving along a MT.
0 denotes an empty site, 1 is K or KT and 2 is KD. Transition from
1 to 2, which corresponds to hydrolysis, occurs within a cell
whereas movement to the forward or backward cell occurs only when
motor is in state 2. At the minus and plus ends the probabilities
are different from those in the bulk.}
\label{fig2}
\end{figure}

Thus, in contrast to the earlier ASEP-like models, each of the self-driven 
particles, which represent the individual motors KIF1A, can be in two 
possible internal states labelled by the indices $1$ and $2$. In other 
words, each of the lattice sites can be in one of three possible allowed 
states (Fig.~\ref{fig2}): empty (denoted by $0$), occupied by a kinesin 
in state $1$, or occupied by a kinesin in state $2$.

For the dynamical evolution of the system, one of the $L$ sites is
picked up randomly and updated according to the rules given below
together with the corresponding probabilities (Fig.~\ref{fig2}):
\begin{eqnarray}
 &&{\rm Attachment:} \,\,\,\,\,  0\to 1 \,\,\,{\rm with} \,\, \omega_a dt\\
 &&{\rm Detachment:} \,\,\,\, 1\to 0 \,\,\, {\rm with} \,\, \ \omega_d dt\\
 &&{\rm Hydrolysis:} \,\,\,\,\,  1\to 2 \,\,\,{\rm with} \,\, \omega_h dt\\
 &&{\rm Ratchet:}\,\,\,\,\, \left\{\begin{array}{c}
     2 \to 1\,\,\,{\rm with} \,\, \omega_s dt\\
     20 \to 01\,\,\,{\rm with} \,\, \omega_f dt
   \end{array}\right.\\
 &&{\rm Brownian\ motion:}\,\,\,\,\, \left\{\begin{array}{c}
     20 \to 02\,\,\,{\rm with} \,\, \omega_b dt\\
     02 \to 20\,\,\,{\rm with} \,\, \omega_b dt
   \end{array}\right.
\end{eqnarray}

The probabilities of detachment and attachment at the two ends of
the MT may be different from those at any bulk site.
We chose $\alpha$ and $\delta$, instead of $\omega_a$, as the
probabilities of attachment at the left and right ends, respectively.
Similarly, we took $\gamma_1$ and $\beta_1$, instead of $\omega_d$,
as probabilities of detachments at the two ends (Fig.~\ref{fig2}). Finally,
$\gamma_2$ and $\beta_2$, instead of $\omega_b$, are the probabilities
of exit of the motors through the two ends by random Brownian movements.
                                                                                
It is possible to relate the rate constants $\omega_f$, $\omega_s$ 
and $\omega_b$
with the corresponding physical processes in the Brownian ratchet
mechanism of a single KIF1A motor. Suppose, just like models of
flashing ratchets \cite{julicher,reimann}, the motor ``sees'' a
time-dependent effective potential which, over each biochemical cycle,
switches back and forth between (i) a periodic but asymmetric sawtooth
like form and (ii) a constant. The rate constant $\omega_h$ in our
model corresponds to the rate of the transition of the potential from
the form (i) to the form (ii). The transition from (i) to (ii) happens
soon after ATP hydrolysis, while the transition from (ii) to (i) happens
when ATP attaches to a bare kinesin\cite{okada1}. The rate constant
$\omega_b$ of the motor in state $2$ captures the Brownian motion of the
free particle subjected to the flat potential (ii). The rate constants
$\omega_s$ and $\omega_f$ are proportional to the overlaps of the
Gaussian probability distribution of the free Brownian particle with,
respectively, the original well and the well immediately in front of the
original well of the sawtooth potential.

Good estimates for the parameters of the model could be extracted by 
analyzing the empirical data \cite{nishietal}. For example, 
$\omega_d \simeq 0.0001$ ms$^{-1}$ is independent of the kinesin 
concentration. On the other hand, $\omega_a$, which depends on the 
kinesin concentration, could be in the range 
$0.0001$ ms$^{-1} \leq \omega_a \leq 0.01$ ms$^{-1}$. Similarly, 
$\omega_b \simeq 0.6$~ms$^{-1}$, $\omega_s \simeq 0.145$~ms$^{-1}$, 
$\omega_f \simeq 0.055$~ms$^{-1}$ and $0 \leq \omega_h \leq 0.25$~ms$^{-1}$.
                                                                                
Let us denote the probabilities of finding a KIF1A molecule in
the states $1$ and $2$ at the lattice site $i$ at time $t$ by the
symbols $r_i$ and $h_i$, respectively. In mean-field approximation the
master equations for the dynamics of motors in the bulk of the system
are given by
\begin{eqnarray}
\frac{dr_i}{dt}&=&\omega_a (1-r_i-h_i) -\omega_h r_i -\omega_d r_i
+\omega_s h_i
+\omega_f h_{i-1}(1-r_i-h_i),\\
\frac{dh_i}{dt}&=&-\omega_s h_i +\omega_h r_i
-\omega_f h_i (1-r_{i+1}-h_{i+1}) \nonumber\\
&&-\omega_b h_i (2-r_{i+1}-h_{i+1}-r_{i-1}-h_{i-1})  
+\omega_b (h_{i-1}+h_{i+1})(1-r_i-h_i).
\label{eq-bulk}
\end{eqnarray}
The corresponding equations for the boundaries are also similar 
\cite{unpub}.

\begin{table}
\begin{tabular}{|c|c|c|c|c|} \hline
ATP (mM)&  $\omega_h$ (1/ms)& $v$ (nm/ms)&  $D/v$ (nm) & $\tau$ (s)\\\hline
$\infty$ & 0.25 & 0.201 & 184.8 & 7.22 \\ \hline
0.9      & 0.20 & 0.176 & 179.1 & 6.94 \\ \hline
0.3375   & 0.15 & 0.153 & 188.2 & 6.98 \\ \hline
0.15     & 0.10 & 0.124 & 178.7 & 6.62 \\ \hline
\end{tabular}
\caption{\label{tab-1mol}{Predicted transport properties
from our model \cite{nishietal} in the low-density limit for 
four different ATP concentrations. $\tau$ is calculated by 
averaging the intervals between attachment and detachment of each KIF1A.}}
\end{table}
\begin{figure}[htb]
\begin{center}
\includegraphics[width=0.9\columnwidth]{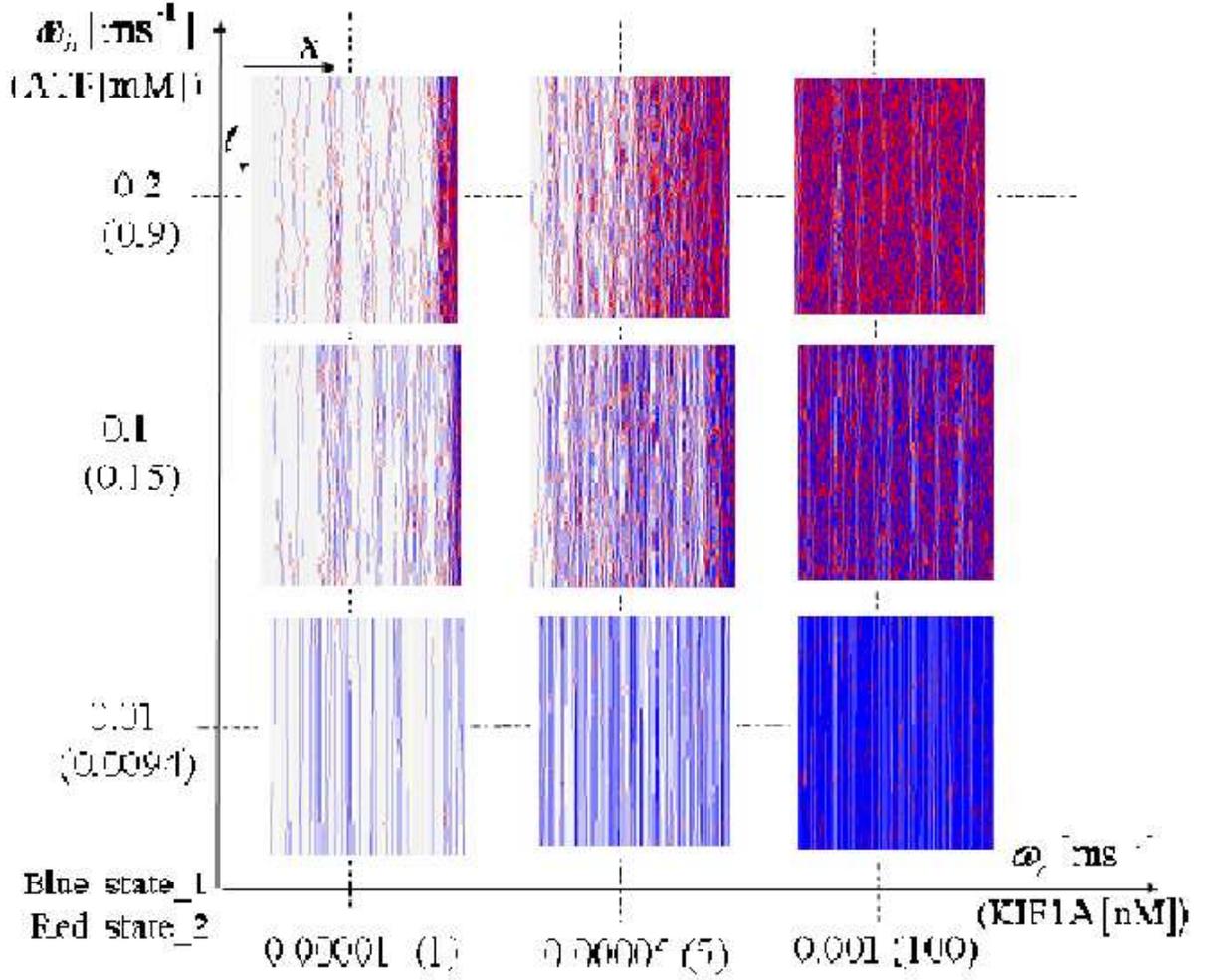}
\end{center}
\caption{Phase diagram of our model \cite{nishietal} in the 
$\omega_h-\omega_a$ plane, with the corresponding values for 
ATP and KIF1A concentrations given in brackets.  These quantities 
are controllable in experiment. The boundary rates are $\alpha=\omega_a,
 \beta_{1,2}=\omega_d, \gamma_{1,2}=\delta=0$.
The position of the immobile shock depends on both ATP and KIF1A
concentrations.
}
\label{fig-kifdiag}
\end{figure}

Assuming that each time step of updating corresponds to 1 ms of real 
time, we performed simulations upto 1 minute. In the limit of low 
density of the motors we have computed, for example, the mean speed of 
the kinesins, the diffusion constant and mean duration of the movement 
of a kinesin on a microtubule from simulations of our model (see table 
\ref{tab-1mol}); these are  in excellent {\it quantitative} agreement 
with the corresponding empirical data from single molecule experiments.

Using this model we have also calculated the flux of the motors in the 
mean field approximation imposing periodic boundary conditions. 
Although the system with periodic boundary conditions is fictitious,
the results provide good estimates of the density and flux in the
corresponding system with open boundary conditions. 

In contrast to the phase diagrams in the $\alpha-\beta$-plane reported
by earlier investigators \cite{frey,santen1,lipo7}, we have drawn the
phase diagram of our model in the $\omega_a-\omega_h$ plane (see 
fig.\ref{fig-kifdiag}) by carrying out extensive computer simulations
for realistic parameter values of the model with open boundary conditions.
The phase diagram shows the strong influence of hydrolysis on the 
spatial distribution of the motors along the MT. For very low
$\omega_h$ no kinesins can exist in state 2; the kinesins, all of which
are in state 1, are distributed rather homogeneously over the entire
system. In this case the only dynamics present is due to the Langmuir
kinetics. At a small, but finite, rate $\omega_h$ both the density 
profiles $\rho_j^{1}$ and $\rho_j^{2}$ of kinesins in the states 1 and 
2 exhibit a localized shock. Interestingly, the shocks in these two 
density profiles always appear at the {\em same} position. Moreover, 
the position of the immobile shock depends on the concentration of the 
motors as well as that of ATP; the shock moves towards the minus end of 
the MT with the increase of the concentration of kinesin or ATP or both 
(Fig.~\ref{fig-kifdiag}).

\begin{figure}[t]
\begin{center}
\vspace{0.5cm}
\includegraphics[width=0.7\columnwidth]{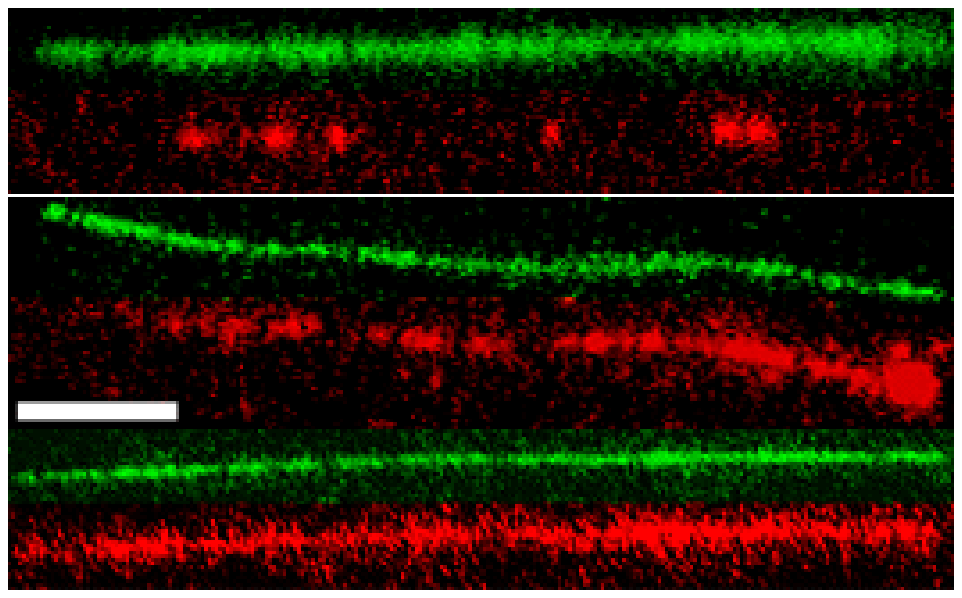}
\end{center}
\caption{Formation of comet-like accumulation of kinesin at the end of
 MT.  Fluorescently labeled KIF1A (red) was introduced to MT (green)
at 10 pM (top), 100 pM (middle) and 1000 pM (bottom) concentrations
along with 2 mM ATP.  The length of the white bar is 2$\mu m$.
}
\label{fig-comet}
\end{figure}

The formation of the shock has been established by our direct 
experimental evidence. The ``comet-like structure'', shown in the
middle of Fig.~\ref{fig-comet}, is the collective pattern formed by
the red fluorescent labelled kinesins where a domain wall separates
the low-density region from the high-density region. The position of
the domain wall depends on both ATP and KIF1A concentrations. Our 
findings on the domain wall are in qualitative agreement with the 
corresponding experimental observations.

\section{Extracellular transport: collagen-based motors}
\label{sec-extra}

The extracellular matrix (ECM) \cite{nagase} of vertebrates is rich in 
collagen. Monomers of collagen form a triple-helical structure which 
self-assemble into a tightly packed periodic organization of fibrils. 
Cells residing in tissues can secret matrix metalloproteases (MMPs), 
a special type of enzymes that are capable of degrading macromolecular 
constituents of the ECM. The most notable among these enzymes is MMP-1 
that is known to degrade collagen. The collagen fibril contains cleavage 
sites which are spaced at regular intervals of $300$ nm. The collagenase 
MMP-1 cleaves all the three $\alpha$ chains of the collagen monomer at 
a single site.

\begin{figure}[t]
\begin{center}
\vspace{0.5cm}
\includegraphics[angle=-90,width=0.7\columnwidth]{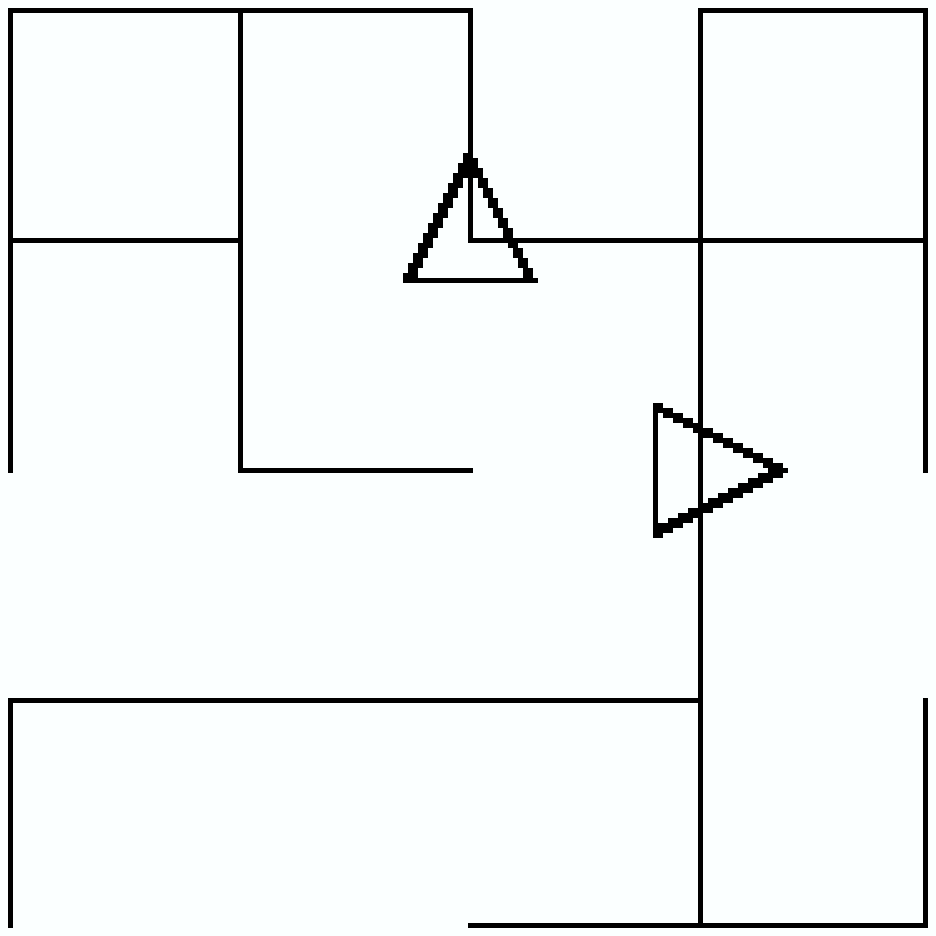}
\end{center}
\caption{Schematic representation of the two-dimensional burnt bridge 
model of MMP-1 dynamics. Each triangle represents a MMP-1 and the 
erased segments represent the ``degraded'' segments of the collagen 
fibril.
}
\label{fig-burnt2}
\end{figure}

Breakdown of the ECM forms an essential step in several biological 
processes like, for example, embryonic development, tissue remodelling, 
etc. \cite{nagase}. Malfunctioning of MMP-1 has been associated with 
wide range of diseases \cite{whittaker}. Therefore, an understanding 
of the MMP-1 traffic on collagen fibrils can provide deeper insight 
into the mechanism of its operation which, in turn, may give some clue 
as to the strategies of control and cure of diseases caused by the 
inappropriate functions of these enzymes.

\subsection{Phenomenology of MMP-1 dynamics} 

Saffarian et al. \cite{saffarian} used a technique of two-photon 
excitation fluorescence correlation spectroscopy to measure the 
correlation function corresponding to the MMP-1 moving along the 
collagen fibrils. The measured correlation function strongly 
indicated that the motion of the MMP-1 was not purely diffusive, 
but a combination of diffusion and drift. In other words, the 
``digestion'' of a collagen fibril occurs when a MMP-1 executes a 
biased diffusion processively (i.e., without detachment) along 
the fibril. They also demonstrated that inactivation of the 
enzyme eliminates the bias but the diffusion remains practically 
unaffected. They claimed that the energy required for the active 
motor-like transport of the MMP-1 comes from the proteolysis (i.e., 
degradation) of the collagen fibrils.

\begin{figure}[t]
\begin{center}
\vspace{0.5cm}
\includegraphics[angle=-90,width=0.7\columnwidth]{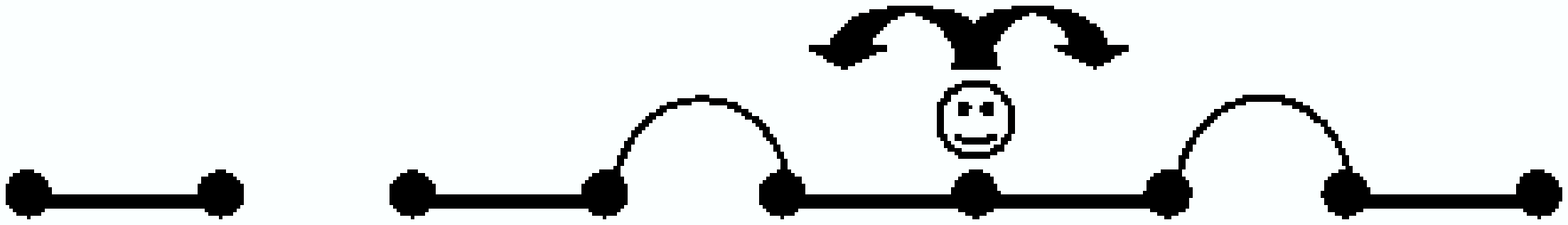}
\end{center}
\caption{Schematic representation of the one-dimensional burnt bridge 
model of MMP-1 dynamics proposed in ref.\cite{blumen}. The thick 
straight lines cannot be burnt whereas the thin arches can get burnt 
by the random walker (see the text for details of the dynamics). The 
erased links represent the burnt bridges.
}
\label{fig-burnt1}
\end{figure}

Saffarian et al.\ \cite{saffarian} also carried out computer 
simulations of a two-dimensional model of the MMP-1 dynamics on 
collagen fibrils; this model is a generalization of the one-dimensional 
``burnt bridge model'' introduced earlier by Mai et al. \cite{blumen}. 
(We shall discuss this model in the next subsection). By comparing 
the results of these simulations with their experimental observations, 
Saffarian et al. they concluded that the observed biased diffusion 
of the MMP-1 on collagen fibrils can be described quite well by a 
Brownian ratchet mechanism \cite{julicher,reimann}.

\subsection{A stochastic burnt-bridge model of MMP-1 dynamics} 

A one-dimensional ``burnt-bridge'' model was developed by Mai et al. 
\cite{blumen} to demonstrate a novel mechanism of directed transport 
of a Brownian particle. 
The model is sketched schematically in fig.\ref{fig-burnt1}. A 
``particle'' performs a random walk on a {\it semi-infinite} 
{\it one-dimensional} lattice that extends from the origin to 
$+\infty$. Each site of the lattice is connected to the two nearest 
neighbour sites by links; a fraction $c$ of these links are called 
``bridges'' and these are prone to be burnt by the random walker. 
A bridge is burnt, with probability $p$, if the random walker either 
crosses it {\it from left to right} or {\it attempts to cross if 
from right to left} \cite{blumen,antal}. In either case, if the 
bridge is actually burnt, 
the walker stays on the right of the burnt bridge and cannot cross 
it any more. The hindrance against leftward motion, that is created 
by the burnt bridges, is responsible for the overall rightward drift 
of the random walker.

Mai et al.\cite{blumen} studied the dependence of the average drift 
velocity $v$ on the parameters $p$ and $c$ by computer simulation. 
They also derived approximate analytical forms of these dependences 
in the two limits $p \ll 1$ and $p \simeq 1$ using a continuum 
approximation. 

Almost every event of crossing of a bridge from left to right or attempt 
of crossing from right to left burns the bridge in the limit $p \simeq 1$.
Whenever the walker burns a bridge, it can take the right edge of the 
burnt bridge as the new origin. Thus, every event of burning of a bridge 
defines a new segment of the lattice having a burnt bridge at its left 
end and a intact bridge at its right end. The random walker performs 
its diffusive movement in the segment such that there is a reflecting 
boundary for the random walker at the left end of the segment and an 
absorbing boundary at the right end of the segment. If the bridges 
were equispaced, each of these segments will have a length $\ell = 1/c$. 
Therefore, $\tau$, the time taken to cross the distance $\ell$ will be 
given by $\ell^{2} = 2 D \tau$ and, hence, the corresponding speed of 
the walker $v = \ell/\tau = 2 D/\ell = 2 c D$. Mai et al. argued that 
if the bridges are intially distributed randomly, the average speed 
will be reduced to $v = c D$. Thus, in the limit $p \rightarrow 1$, 
$v \propto c$.  In contrast, in the limit $p \ll 1$, they found 
$v \propto \sqrt{pc}$. 

\section{Cellular traffic} 
\label{sec-cellular}

A {\it Mycoplasma mobile} (MB) bacterium is an uni-cellular organism. 
Each of the pear-shaped cells of this bacterium is about $700$ nm long 
and has a diameter of about $250$ nm at the widest section. Each 
bacterium can move fast on glass or plastic surfaces using a {\it 
gliding} mechanism. 

\begin{figure}[ht]
\begin{center}
\includegraphics[angle=-90,width=0.48\columnwidth]{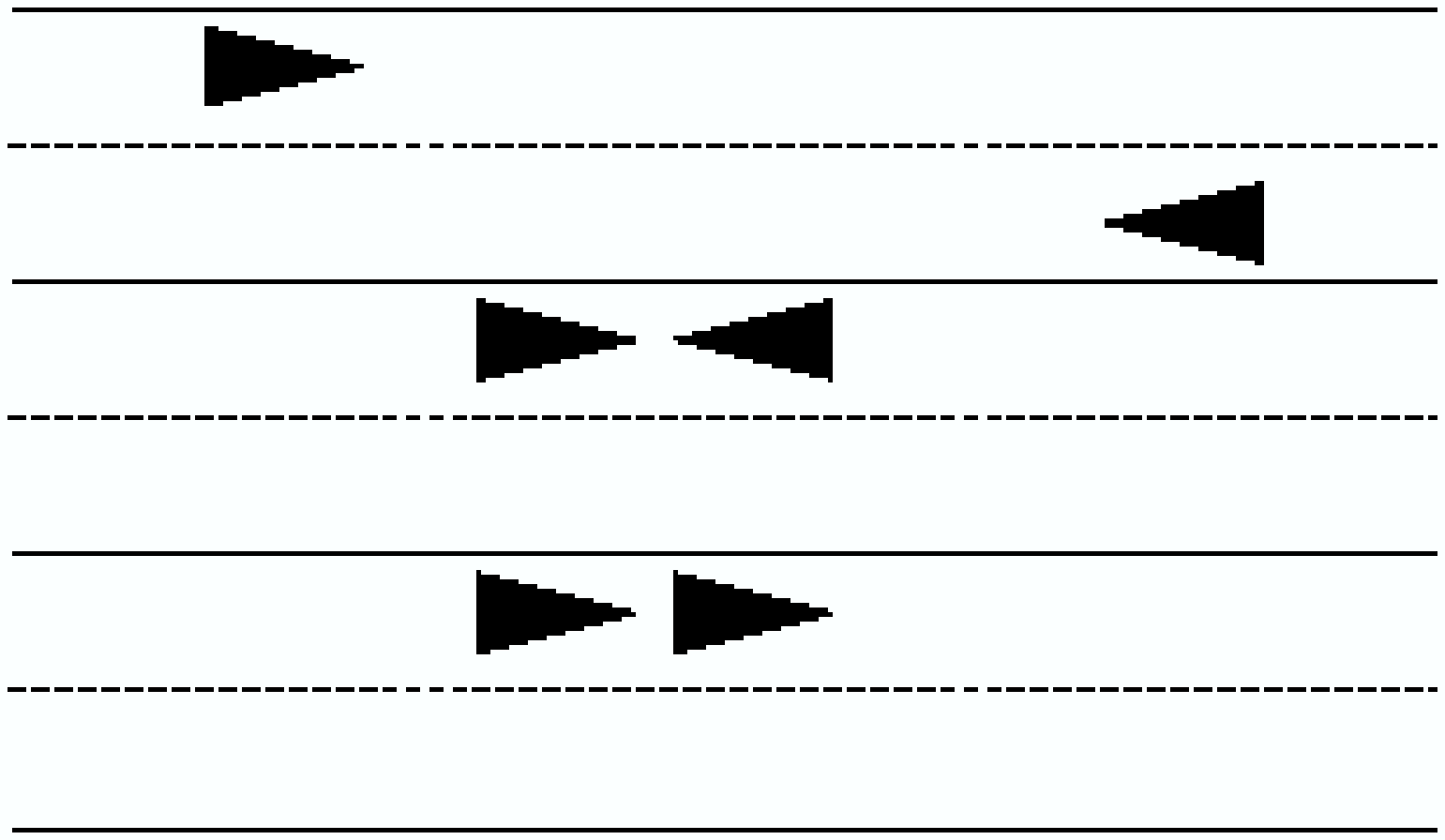}
\includegraphics[angle=-90,width=0.48\columnwidth]{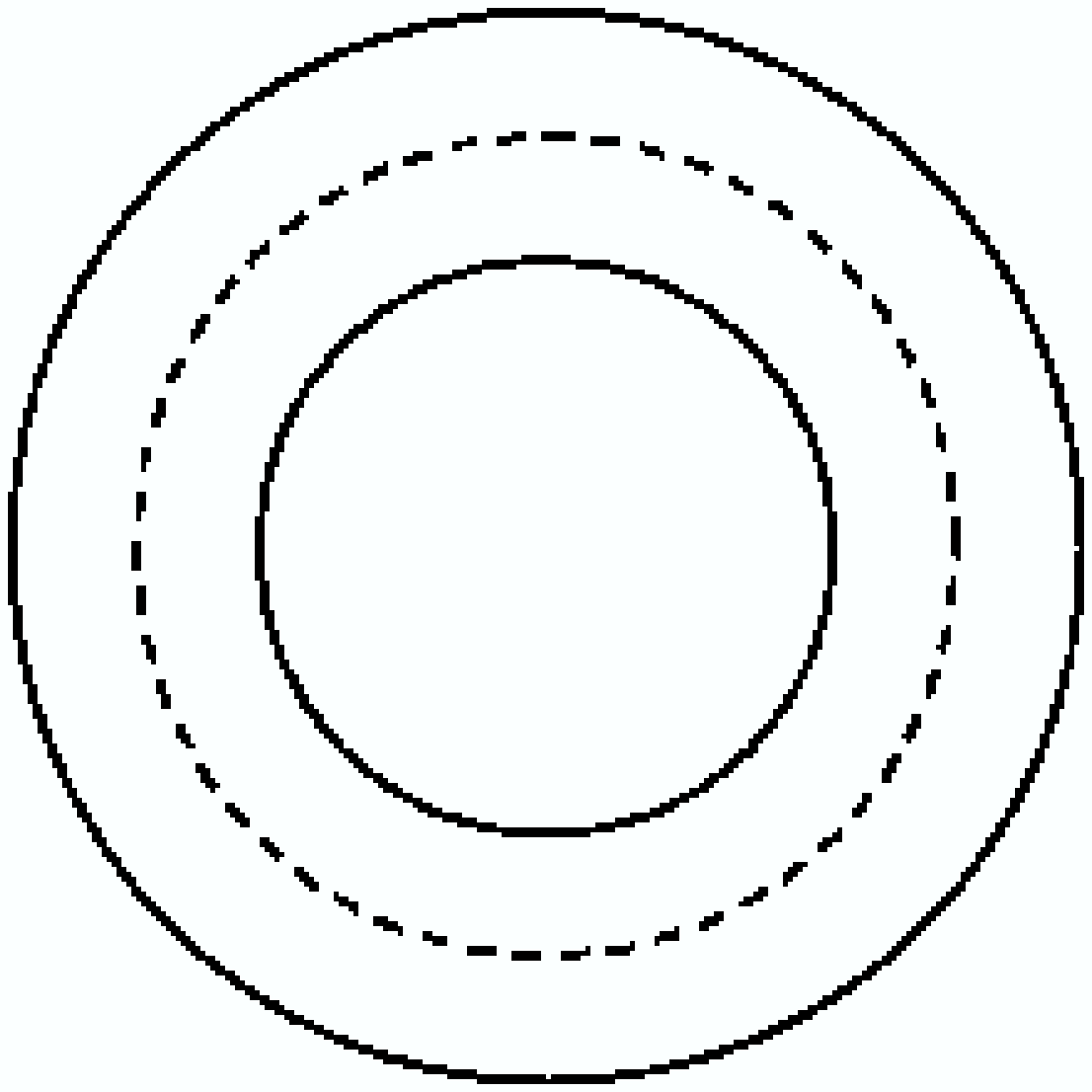}
\end{center}
\caption{Schematic description of the traffic flow of {\it Mycoplasma} 
(represented by the triangles) in (a) the expertment of Hiratsuka et al. 
(b) the single-channel model of Vartak et al. \cite{vartak}. The width 
of each channel in both (a) and (b) is just enough to allow the passage 
of two cells side by side. Both uni-directional and bi-directional 
traffic are possible.
}
\label{fig-density}
\end{figure}

In a recent experiment \cite{hiratsuka} narrow linear channels were 
constructed on lithographic substrates. The channels were typically 
$500$ nm wide and $800$ nm deep. Note that each chennel was approximately 
twice as wide as the width of a single MB cell (see the sketch on the 
left side of the fig.\ref{fig-density}). The channels were so deep that 
none of the individual MB cells was able to climb up the tall walls of 
the channels and continued moving along the bottom edge of the walls 
of the cannels. In the absence of direct contact interaction with other 
bacteria, each individual MB cell was observed to glide, without changing 
direction, at an average speed of a few microns per second. 

When two MB cells made a contact approaching each other from opposite 
directions within the same channel, one of the two cells gave way and 
moved to the adjacent lane. However, in a majority of the cases, two 
cells approaching each other from the opposite directions simply passed 
by as if nothing had happened; this is because of the fact that the 
width of the chennel is roughly twice that of the individual MB cell.  
Moreover, when two cells moving in the same direction within a channel 
collided with each other, the faster cell moved to the adjacent lane 
after the collision. 

Hiratsuka et al.\cite{hiratsuka} attached micron-sized beads on the 
MB cells using biochemical technique and demonstrated that the average 
speed of each MB cell remained practically unaffected by the load it 
was carrying. In contrast to the nonliving motile elements discussed 
in all the preceedings sections, the cells are the functional units of 
life. Therefore, the MB cells have the potential for use in applied 
research and technology as ``micro-transporters''. More recently, the 
unicellular biflagellated algae {\it Chlamydomonas reinhardtii} (CR), 
which are known to be phototactic {\it swimmers}, have been shown to 
be even better cadidates as ``micro-transporter'' as these can attain 
average speeds that is about two orders of magnitude higher than what 
was possible with MB cells \cite{weibel}. However, to our knowledge, 
the effects of mutual interactions of the CR cells on their average 
speed at higher densities has not been investigated.

\section{Traffic in social insect colonies: ants and termites}
\label{sec-ants}

From now onwards, in this review we shall study traffic of 
multi-cellular organisms. We begin with the simpler (and smaller) 
organisms and, then, consider those of organisms with larger sizes 
and more complex physiology in the next section.

Termites, ants, bees and wasps are the most common social insects,
although the extent of social behavior, as compared to solitary
life, varies from one sub-species to another \cite{wilson}. The
ability of the social insect colonies to function without a leader
has attracted the attention of experts from different disciplines
\cite{bonabu97,anderson02,huang,bonabu98,theraulaz03,gautrais,
keshet94,theraulazetal}. Insights gained from the modeling of the
colonies of such insects are finding important applications
in computer science (useful optimization and control algorithms)
\cite{dorigo}, communication engineering \cite{bona00}, artificial
``swarm intelligence'' \cite{bonabeau} and micro-robotics \cite{krieger}
as well as in task partitioning, decentralized manufacturing
\cite{anderson99a,anderson99b,anderson99c,anderson00a,anderson01,anderson00b}
and management \cite{meyer}.

In this section we consider only ants as the collective terrestrial movements 
of these have close similarities with the other traffic-like phenomena 
considered here. When observed from a sufficiently long distance the 
movement of ants on trails resemble the vehicular traffic observed from 
a low flying aircraft.

Ants communicate with each other by dropping a chemical (generically
called {\it pheromone}) on the substrate as they move forward
\cite{wilson,camazine,mikhailov}. Although we cannot smell it, the
trail pheromone sticks to the substrate long enough for the other
following sniffing ants to pick up its smell and follow the trail.
This process is called {\em chemotaxis} \cite{wilson}.

Rauch et al.\cite{rauch} developed a continuum model, following a
hybrid of the individual-based and population-based approaches in terms 
of an effective energy landscape. They wrote one set of stochastic
differential equations for the positions of the ants and another set 
of PDEs for the local densities of pheromone. Both this model and the 
CA model introduced by Watmough and Edelstein-Keshet \cite{watmough} 
were intended to address the question of formation of the ant-trail 
networks by foraging ants. 

Couzin and Franks \cite{couzin} developed an individual based model 
that not only addressed the question of self-organized lane formation 
but also elucidated the variation of the flux of the ants with two 
important parameters of the model. The ``internal angle'' $\alpha$ may 
be interpreted as angle of local visual field or that of olfactory 
perception, or tactile range of the antennae of the individual ants. 
Moreover, each individual ant is assumed to turn away from others 
within these zones by, at most, an angle $\theta_a \Delta t$ in time 
$\Delta t$. 

Imposing periodic boundary conditions, Couzin and Franks \cite{couzin} 
computed the flux of ants in their model by computer simulations. The 
flux was found to be a nonmonotinic function of both $\alpha$ and 
$\theta_a$. At low $\alpha$, ants cannot detect others ahead whereas at 
high $\alpha$ they spend most of their time avoiding others even through 
collisions with others may be unlikely; both these reduce the flux 
considerably. Similarly, at low $\theta_a$ ants cannot turn sufficiently 
rapidly to avoid collision whereas at high $\theta$ they change their 
direction quickly so that not many move in the same direction at any 
time. Thus, only in the intermediate range of values of $\alpha$ and 
$\theta_a$, the ants are optimally sensitive. Therefore, the flux 
exhibits a maximum both as a function of $\alpha$ and as a function of 
$\theta_a$.

In the recent years, we have developed discrete models that are not
intended to address the question of the emergence of the ant-trail
\cite{activewalker}, but focus on the traffic of ants on a trail which
has already been formed. We have developed models of both unidirectional 
and bidirectional ant-traffic by generalizing the totally asymmetric
simple exclusion process (TASEP) \cite{derrida1,derrida2,gunterrev} with
parallel dynamics by taking into account the effect of the pheromone.

\subsection{Model of single-lane uni-directional ant-traffic} 

\begin{figure}[ht]
\begin{center}
\includegraphics[width=0.65\textwidth]{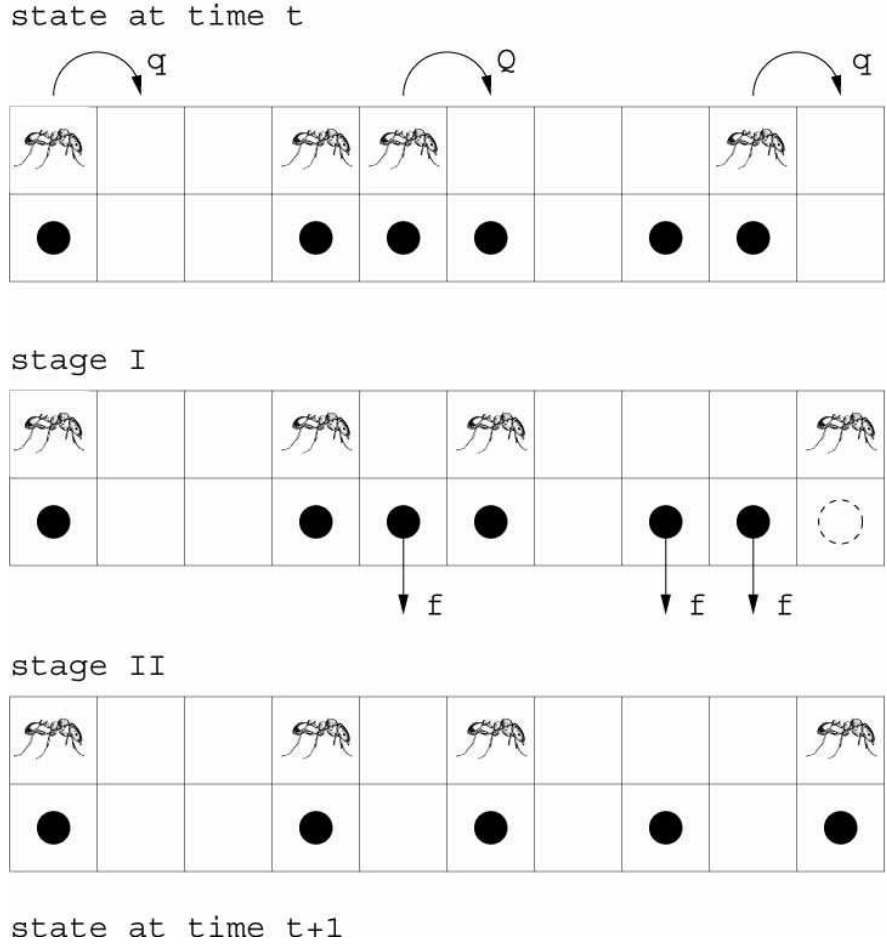}
\end{center}
\caption{
Schematic representation of typical configurations of the
uni-directional ant-traffic model. The symbols $\bullet$ indicate
the presence of pheromone.
This figure also illustrates the update procedure.
Top: Configuration at time $t$, i.e.\ {\it before} {\em stage I}
of the update. The non-vanishing probabilities of forward movement of
the ants are also shown explicitly. Middle: Configuration {\it after}
one possible realisation of {\it stage I}. Two ants have moved compared
to the top part of the figure. The open circle with dashed boundary
indicates the location where pheromone will be dropped by the corresponding
ant at {\em stage II} of the update scheme. Also indicated are the existing
pheromones that may evaporate in {\em stage II} of the updating, together
with the average rate of evaporation.  Bottom: Configuration {\it after}
one possible realization of {\it stage II}. Two drops of pheromones
have evaporated and pheromones have been dropped/reinforced at the
current locations of the ants.
}
\label{fig-modeldef1}
\end{figure}

In our model of uni-directional ant-traffic the ants move according to
a rule which is essentially an extension of the TASEP dynamics.
In addition, a second field is introduced which models
the presence or absence of pheromones (see Fig.~\ref{fig-modeldef1}).
The hopping probability of the ants is now modified by the presence of
pheromones. It is larger if a pheromone is present at the destination site.
Furthermore, the dynamics of the pheromones has to be specified. They
are created by ants and free pheromones evaporate with probability $f$
per unit time. Assuming periodic boundary conditions, the state of the
system is updated at each time step in two stages
(see Fig.~\ref{fig-modeldef1}). In stage I ants are allowed to move
while in stage II the pheromones are allowed to evaporate. In each
stage the {\it stochastic} dynamical rules are applied in parallel to
all ants and pheromones, respectively.\\

\begin{figure}[ht]
\begin{center}
\centerline{(a)}
\vspace{.5cm}
\includegraphics[width=0.65\textwidth]{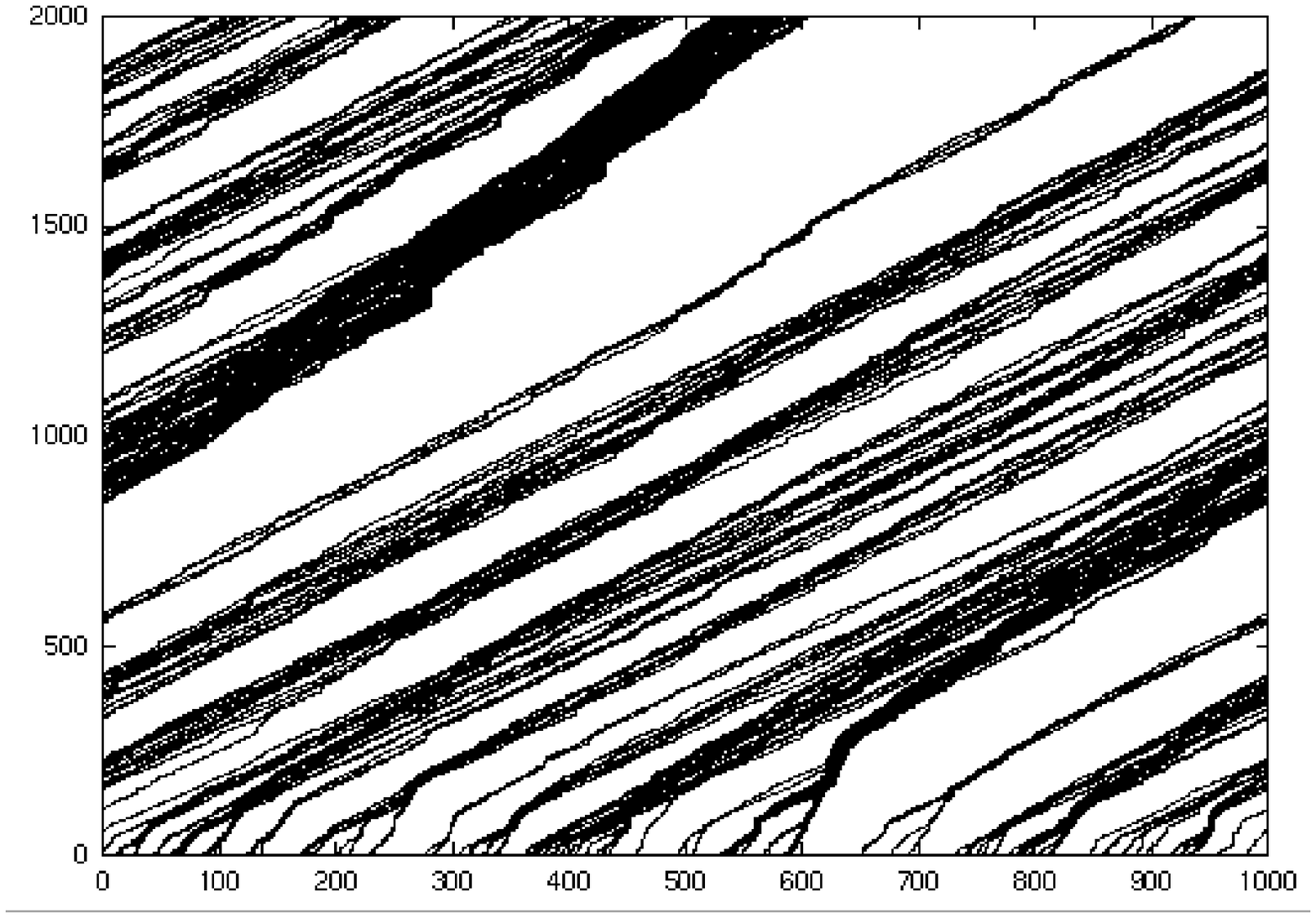}
\vspace{.5cm}
\centerline{(b)}
\vspace{.5cm}
\includegraphics[width=0.65\textwidth]{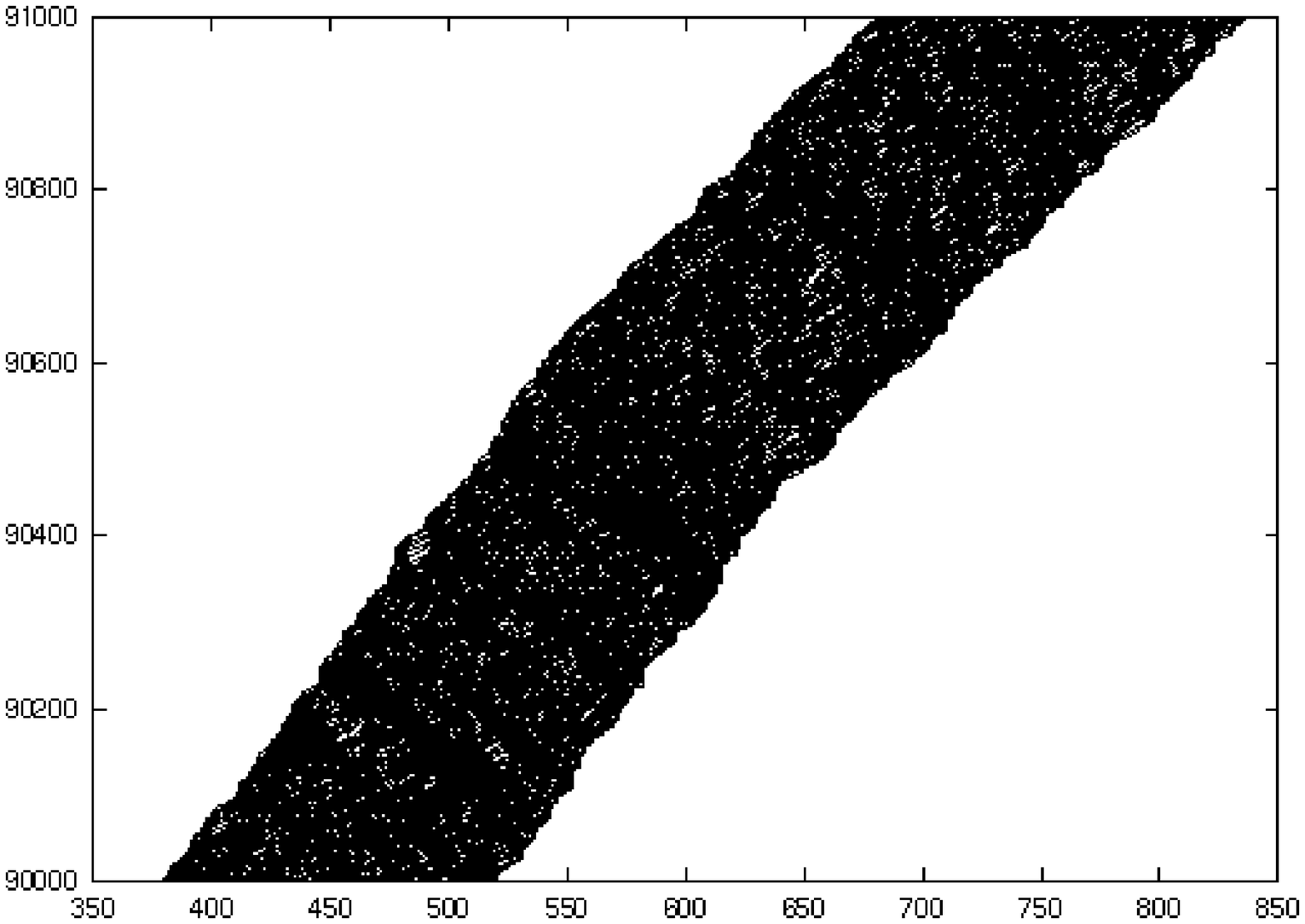}
\end{center}
\caption{
Snapshots of the spatial configurations demonstrating coarsening of
the clusters of ants (a) in the early stages and (b) in the late 
stages of time evolution starting from random initial condition. The 
position of each ant is denoted by a black dot. A horizontal row of 
dots on such {\it space-time plots} denotes the positions of the ants 
(i.e., a snapshot of the system) at a given instant of time; an upper 
row corresponds to the snapshot at a later time. 
}
\label{fig-coarse}
\end{figure}

\noindent {\it Stage I: Motion of ants}\\[0.2cm]
\noindent An ant in a site cannot move if the site immediately in front
of it is also occupied by another ant. However, when this site is not
occupied by any other ant, the probability of its forward movement to
the ant-free site is $Q$ or $q$, depending on whether or not the target
site contains pheromone. Thus, $q$ (or $Q$) would be the average speed
of a {\it free} ant in the absence (or presence) of pheromone. To be
consistent with real ant-trails, we assume $ q < Q$, as presence of
pheromone increases the average speed.\\
                                                                                
\noindent {\it Stage II: Evaporation of pheromones}\\[0.2cm]
\noindent Trail pheromone is volatile. So, pheromone secreted by an ant
will gradually decay unless reinforced by the following ants. In order to
capture this process, we assume that each site occupied by an ant at the
end of stage I also contains pheromone. On the other hand, pheromone in
any `ant-free' site is allowed to evaporate; this evaporation is also
assumed to be a random process that takes place at an average rate of $f$
per unit time.\\
                                                                                
The total amount of pheromone on the trail can fluctuate although the
total number $N$ of the ants is constant because of the
periodic boundary conditions. In the two special cases $f = 0$ and
$f = 1$ the stationary state of the model becomes identical to that of
the TASEP with hopping probability $Q$ and $q$, respectively.
                                                                                
One interesting phenomenon observed in the simulations is coarsening.
At intermediate time usually several non-compact clusters are formed
(Fig.~\ref{fig-coarse}(a)). However, the velocity of a cluster depends
on the distance to the next cluster ahead. Obviously, the probability
that the pheromone created by the last ant of the previous cluster
survives decreases with increasing distance. Therefore clusters with
a small headway move faster than those with a large headway.
This induces a coarsening process such that after long times only
one non-compact cluster survives (Fig.~\ref{fig-coarse}(b)). A similar
behaviour has been observed also in the bus-route model {\footnote{ 
In the bus route model, each bus stop can accomodate at most one bus 
at a time; the passengers arrive at the bus stops randomly at an 
average rate $\lambda$ and each bus, which normally moves from one 
stop to the next at an average rate $Q$, slows down to $q$, to pick 
up waiting passengers }} \cite{loan,cd}. If the system evolves from 
a random initial condition at $t = 0$, then during coarsening of the 
cluster, its size $R(t)$ at time $t$ is given by $R(t) \sim t^{1/2}$ 
\cite{loan,cd}. 

\vspace{1cm}

\begin{figure}[ht]
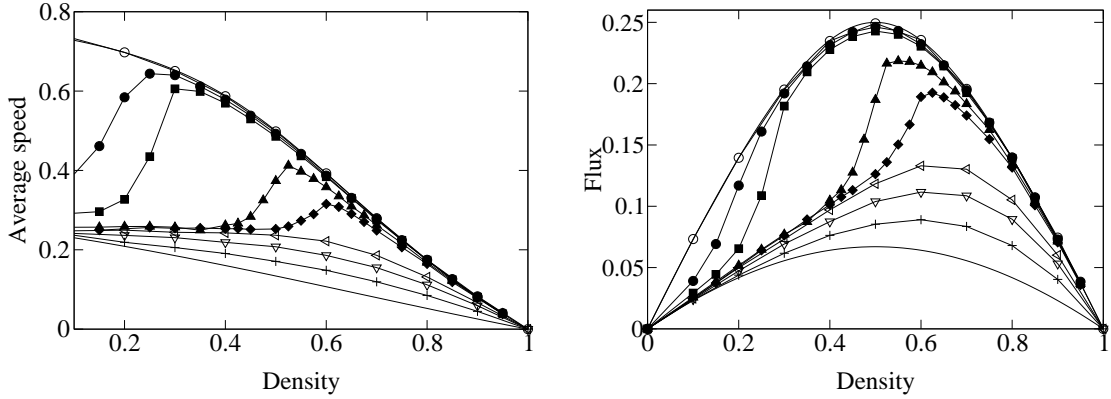

\begin{center}
\centerline{\epsfig{figure=fig22a.eps,width=7cm}\qquad
\epsfig{figure=fig22b.eps,width=7cm}}
\end{center}
\caption{ The variation of the average speed (left) and flux (right) of 
the ants with their density on trail. 
}
\label{fig-antspeed}
\end{figure}

In vehicular traffic, usually, the inter-vehicle interactions tend to
hinder each other's motion so that the average speed of the vehicles
decreases {\it monotonically} with increasing density. In contrast,
in our model of uni-directional ant-traffic the average speed of the
ants varies {\it non-monotonically} with their density over a wide
range of small values of $f$ because of the coupling of their dynamics
with that of the pheromone (see fig.\ref{fig-antspeed}). This uncommon 
variation of the average speed gives rise to the unusual dependence of 
the flux on the density of the ants in our uni-directional ant-traffic 
model (see fig.\ref{fig-antspeed}). Furthermore, the flux is no longer 
particle-hole symmetric.

\subsection{Model of single-lane bi-directional ant-traffic} 

\begin{figure}[ht] 
\begin{center} 
\includegraphics[width=0.5\columnwidth]{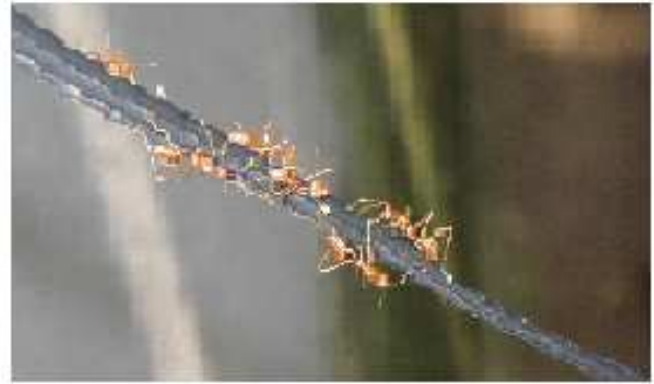} 
\end{center} 
\caption{A snapshot of an ant-trail on a hanging cable. It can be 
regarded as strictly one-dimensional. But, nevertheless, traffic 
flow in opposite directions is possible as two ants, which face each 
other on the upper side of the cable, can exchange their positions 
if one of them, at least temporarily, switches over to the lower 
side of the cable.} 
\label{fig-antphoto}
\end{figure}

The single-lane model of uni-directional ant traffic, which we have 
discussed above, has been extended \cite{kunwar} to capture the essential 
features of bi-directional ant-traffic in some special situations 
like, for example, on hanging cables (see fig.\ref{fig-antphoto}).

\begin{figure}
\begin{tabular}{| c | c | c |}
\hline
initial & final & rate\\
\hline
RL & RL & $1-K$\\
   & LR & $K$\\
\hline
RP & RP & $(1-f)(1-Q)$\\
   & R0 & $f(1-Q)$\\
   & 0R & $fQ$\\
   & PR & $(1-f)Q$\\
\hline
R0 & R0 & $1-q$\\
   & 0R & $fq$\\
   & PR & $(1-f)q$\\
\hline
\end{tabular}\ \ \
\begin{tabular}{| c | c | c |}
\hline
initial & final & rate\\
\hline
PR & PR & $1-f$\\
   & 0R & $f$\\
\hline
P0 & P0 & $1-f$\\
   & 00 & $f$\\
\hline
PP & PP & $(1-f)^2$\\
   & P0 & $f(1-f)$\\
   & 0P & $f(1-f)$\\
   & 00 & $f^2$\\
\hline
\end{tabular}
\caption{Nontrivial transitions and their transition rates in the PRL 
model \cite{kunwar}. Transitions from initial states $PL$, $0L$ and 
$0P$ are not listed. They can be obtained from those for $LP$, $L0$ 
and $P0$, respectively, by replacing $R\leftrightarrow L$ and, then, 
taking the mirror image.}
\label{fig-updating}
\end{figure}
\noindent

In our model the right-moving (left-moving) particles, represented by 
$R$ ($L$), are never allowed to move towards left (right); these two 
groups of particles are the analogs of the outbound and nest-bound 
ants in a {\it bi-directional} traffic on the same trail. Thus, no 
U-turn is allowed. In addition to the TASEP-like hopping of the 
particles onto the neighboring vacant sites in the respective directions 
of motion, the $R$ and $L$ particles on nearest-neighbour sites and 
facing each other are allowed to exchange their positions, i.e., the 
transition $ RL \overset{K}{\to} ~~LR$ takes place, with the 
probability $K$. This might be considered as a minimal model for the 
motion of ants on a hanging cable as shown in Fig.\ref{fig-antphoto}. 
When a outbound ant and a nest-bound ant face each other on the upper 
side of the cable, they slow down and, eventually, pass each other 
after one of them, at least temporarily, switches over to the 
lower side of the cable. Similar observations have been made 
for normal ant-trails where ants pass each other after turning by a 
small angle to avoid head-on collision \cite{couzin,burd2}. In our 
model, as commonly observed in most real ant-trails, none of the ants 
is allowed to overtake another moving in the same direction.

We now 
introduce a third species of particles, labelled by the letter $P$, 
which are intended to capture the essential features of pheromone. 
The $P$ particles are deposited on the lattice by the $R$ and $L$ 
particles when the latter hop out of a site; an existing $P$ 
particle at a site disappears when a $R$ or $L$ particle arrives 
at the same location. The $P$ particles cannot hop but can {\it 
evaporate}, with a probability $f$ per unit time, independently 
from the lattice. None of the lattice sites can accomodate more 
than one particle at a time. 

The state of the system is updated in a {\it random-sequential} manner. 
Because of the periodic boundary conditions, the densities of the $R$ 
and the $L$ particles are conserved. In contrast, the density of the $P$ 
particles is a non-conserved variable. The distinct initial states and 
the corresponding final states for pairs of nearest-neighbor sites are 
shown in fig.\ref{fig-updating} together with the respective transition 
probabilties.

\begin{figure}[ht]
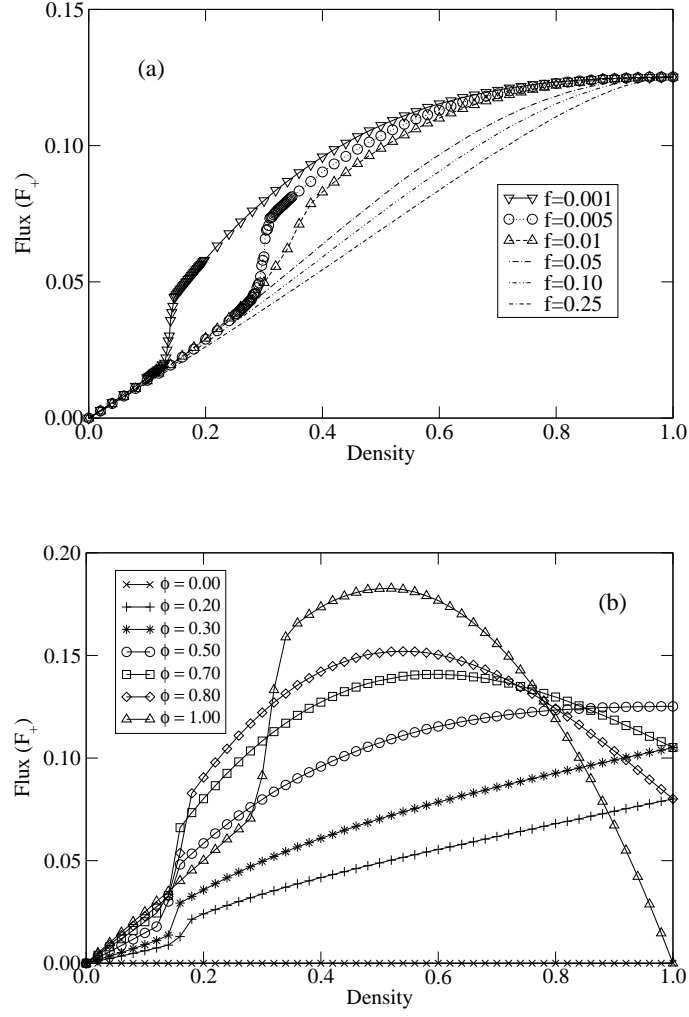

\begin{center}
\includegraphics[width=0.5\textwidth]{fig24a.eps} 
\hspace{1cm}\\
\vspace{1cm}
\includegraphics[width=0.5\textwidth]{fig24b.eps} 
\end{center}
\caption{The fundamental diagrams in the steady-state of the PRL 
model \cite{kunwar} for several different values of (a) $f$ (for 
$\phi = 0.5$) and (b) $\phi$ (for $f = 0.001$). The other common 
parameters are $Q = 0.75, q = 0.25$, $K = 0.5$ and $M = 1000$. 
Non-monotonic variation of the average speeds of the ants with 
their density on the trail gives rise to the unusual shape of the 
fundamental diagrams. 
}
\label{fig-prlfd}
\end{figure}

\begin{figure}[ht]
\begin{center}
\includegraphics[width=0.5\textwidth]{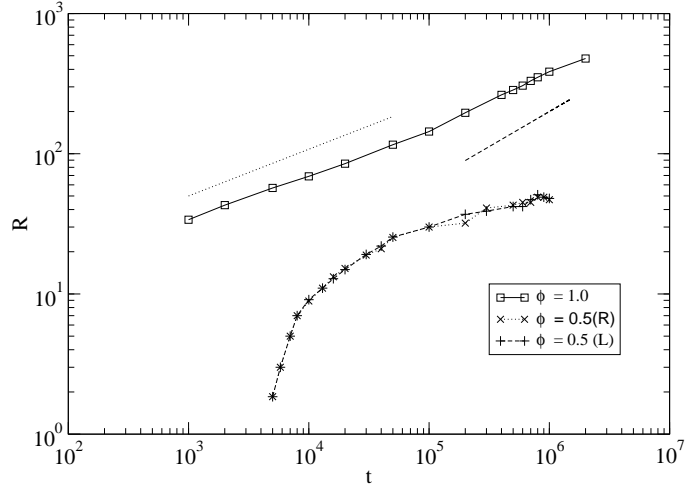} 
\end{center}
\caption{Average size of the cluster $R$ in the PRL model \cite{kunwar} 
is plotted against time $t$ for $\phi = 1.0$, and $\phi = 0.5$, both 
for the same total density $c = 0.2$; the other common parameters being 
$Q = 0.75$, $q = 0.25$, $K = 0.50$, $f = 0.005$, $M = 4000$. }  
\label{fig-rvst}
\end{figure}

Suppose $N_+$ and $N_- = N - N_+$ are the total numbers of $R$ and $L$ 
particles, respectively. For a system of length $M$ the corresponding 
densities are $c_{\pm} = N_{\pm}/M$  with the total density 
$c = c_+ + c_- = N/M$. Of the $N$ particles, a fraction 
$\phi = N_{+}/N = c_{+}/c$ are of the type $R$ while the remaining 
fraction $1-\phi$ are $L$ particles. The corresponding fluxes are 
denoted by $J_{\pm}$. In both the limits $\phi = 1$ and $\phi = 0$ this 
model reduces to the model reported in ref.\cite{cgns,ncs} and reviewed 
in section IX.A, which was motivated by uni-directional ant-traffic and 
is closely related to the bus-route models \cite{loan,cd}.

One unusual feature of this PRL model is that the flux does {\it not} 
vanish in the {\it dense-packing} limit $c \rightarrow 1$. In fact, 
in the {\it full-filling} limit $c = 1$, the {\it exact} non-vanishing 
flux $J_+ = K c_+ c_- = J_-$ at $c_++c_- = c = 1$ arises only from the 
exchange of the $R$ and $L$ particles, {\it irrespective of the 
magnitudes of} $f, Q$ and $q$.

In the special case $Q = q =: q_h$ the hopping of the ants become 
independent of pheromone. This special case of the PRL model is 
identical to the AHR model \cite{arndt} with $q_- = 0 = \kappa$. 
A simple mean-field approximation (MFA) yields the estimates 
\begin{eqnarray}
J_{\pm} \simeq c_{\pm} \biggl[ q_h (1-c) + K c_{\mp} \biggr] 
\label{eq-mf}
\end{eqnarray} 
{\it irrespective of} $f$, for the fluxes $J_{\pm}$ at any arbitrary $c$. 
We found that the results of MFA agree reasonably well with the exact 
values of the flux \cite{rajewsky} for all $q_h \geq 1/2$ but deviate 
more from the exact values for $q_h < 1/2$, indicating the presence of 
stronger correlations at smaller values of $q_h$.

For the generic case $q \neq Q$, the flux in the PRL model depends on 
the evaporation rate $f$ of the $P$ partcles. In Fig.~\ref{fig-prlfd} 
we plot the fundamental diagrams for wide ranges of values of $f$ 
(in Fig.~\ref{fig-prlfd}(a)) and $\phi$ (in Fig.~\ref{fig-prlfd}(b)), 
corresponding to one set of hopping probabilities. 
First, note that the data in figs. \ref{fig-prlfd} are consistent 
with the physically expected value of 
$J_{\pm}(c = 1) = K c_{+} c_{-}$, because in the dense packing 
limit only the exchange of the oppositely moving particles 
contributes to the flux. Moreover, the sharp rise of the flux 
over a narrow range of $c$ observed in both Fig.~\ref{fig-prlfd} (a) 
and (b) arise from the nonmonotonic variation of the average speed 
with density, an effect which was also observed in our earlier model 
for uni-directional ant traffic \cite{cgns,ncs}.

In the special limits $\phi = 0$ and $\phi = 1$, this model reduces 
to our single-lane model of unidirectional ant traffic; therefore, 
in these limits, over a certain regime of density (especially at small 
$f$), the particles are expected \cite{cgns,ncs} to form ``loose'' 
(i.e., non-compact) clusters \cite{ncs}. 
Therefore, in the absence of encounter with oppositely moving particles, 
$\tau_{\pm}$, the coarsening time for the right-moving and left-moving 
particles would grow with system size as $\tau_{+} \sim \phi^2 M^2$ and 
$\tau_{-} \sim (1-\phi)^2 M^2$. 
                                                                               
\begin{figure}[tb]
\begin{center}
\vspace{0.5cm}
\includegraphics[width=0.375\textwidth]{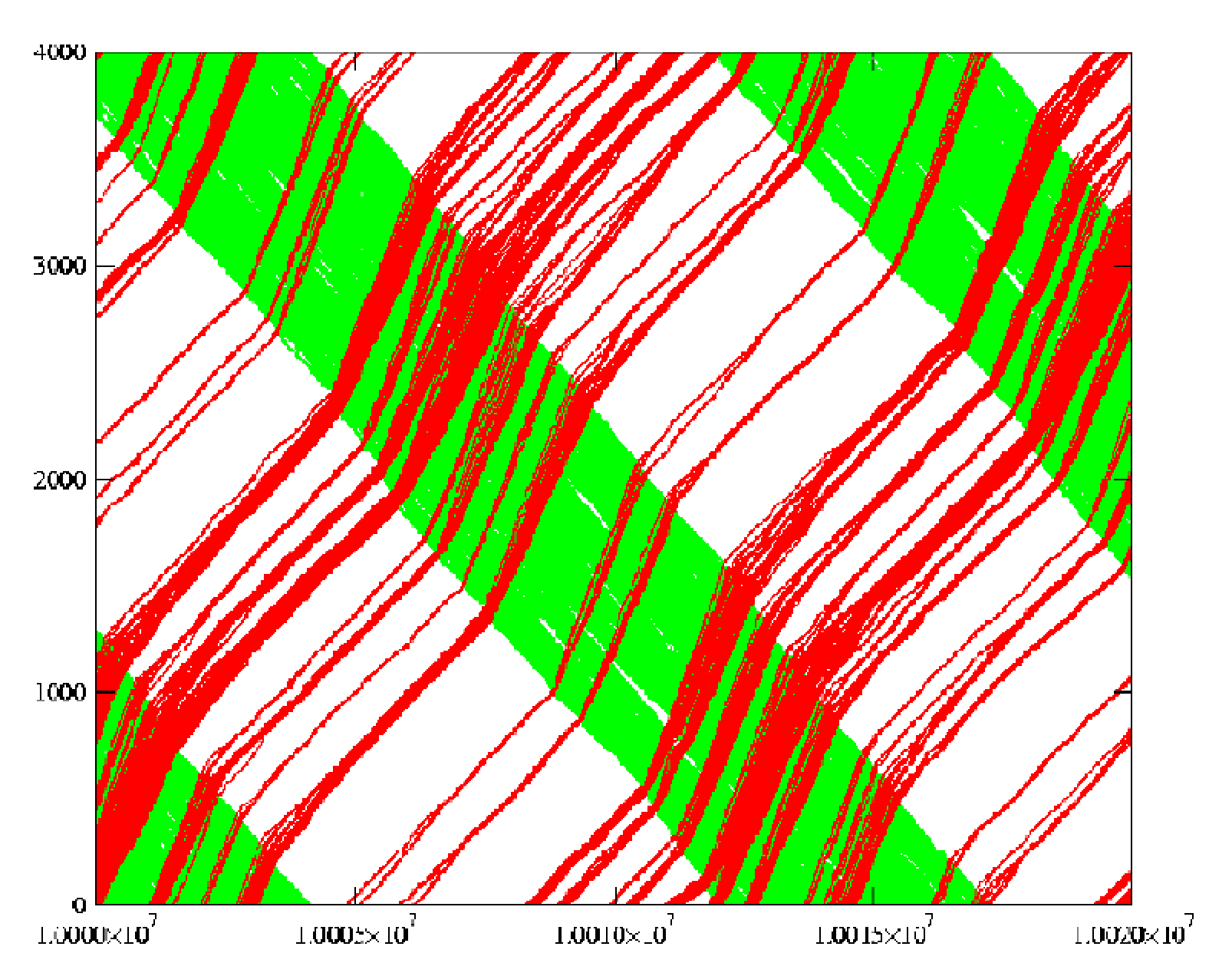}
\end{center}
\caption{Space-time plot of the PRL model for  $Q = 0.50$, $q = 0.25$,
$f = 0.005$, $c = 0.2$, $\phi = 0.3$, $K = 1.0$ and $M = 4000$. The 
black and grey dots represent the right-moving and
left-moving ants, respectively.
}
\label{fig-shred}
\end{figure}

In the PRL model {\it with periodic boundary conditions}, the oppositely 
moving `loose'' clusters ``collide'' against each other periodically 
where the time gap $\tau_g$ between the successive collisions increases 
{\it linearly} with the system size following $\tau_g \sim M$; we have 
verified this scaling relation numerically (see the typical space-time 
diagram in fig.\ref{fig-shred}). 
During a collision each loose cluster ``{\it shreds}'' (i.e., cuts into 
pieces) the oppositely moving cluster; both clusters shred the other 
equally if $\phi = 1/2$. However, for all 
$\phi \neq 1/2$, the minority cluster suffers more severe 
shredding than that suffered by the majority cluster 
 because each member of a cluster 
contributes in the shredding of the oppositely moving cluster. 
In small systems the ``shredded'' clusters get opportunity 
for significant re-coarsening before getting shredded again in 
the next encounter with the oppositely moving particles. But, in 
sufficiently large systems, shredded appearance of the clusters 
persists. However, we observed practically no difference in the 
fundamental diagrams for $M = 1000$ and $M = 4000$.

Following the methods of ref.\cite{cd}, we have computed $R(t)$ 
starting from random initial conditions. The data (Fig.~\ref{fig-rvst}(a)) 
corresponding to $\phi = 1$ are consistent with the asymptotic 
growth law $R(t) \sim t^{1/2}$. In sharp contrast, for $\phi = 0.5$, 
$R(t)$ saturates to a much smaller value (Fig.~\ref{fig-rvst}(b)) that 
is consistent with highly shredded appearance of the corresponding 
clusters.

Thus, coarsening and shredding phenomena compete against each 
other and this competition determines the overall spatio-temporal 
pattern. Therefore, in the late stage of evolution, the system 
settles to a state where, because of alternate occurrence of 
shredding and coarsening, the typical size of the clusters varies 
periodically. Moreover, we find that, for given $c$ and $\phi$, 
increasing $K$ leads to  sharper {\it speeding up} of the clusters 
during collision so long as $K$ is not much smaller than $q$. Both 
the phenomena of shredding and speeding during collisions of the 
oppositely moving loose clusters arise from the fact that, during 
such collisions, the domainant process is the exchange of positions, 
with probability $K$, of oppositely-moving ants that face each other.

\subsection{Model of two-lane bi-directional ant-traffic}
                                                                               
It is possible to extend the model of uni-directional ant-traffic  
to a minimal model of two-lane bi-directional ant-traffic \cite{jscn}.
In such models of bi-directional ant-traffic the trail consists of
{\it two} lanes of sites. These two lanes need not be physically
separate rigid lanes in real space. In the initial configuration, a
randomly selected subset of the ants move in the clockwise direction
in one lane while the others move counterclockwise in the other lane.
The numbers of ants moving in the clockwise direction and
counterclockwise in their respective lanes are fixed, i.e.\ ants are
allowed neither to take U-turn. 
                                                                                
\begin{figure}[ht]
\begin{center}
\includegraphics[width=0.5\textwidth]{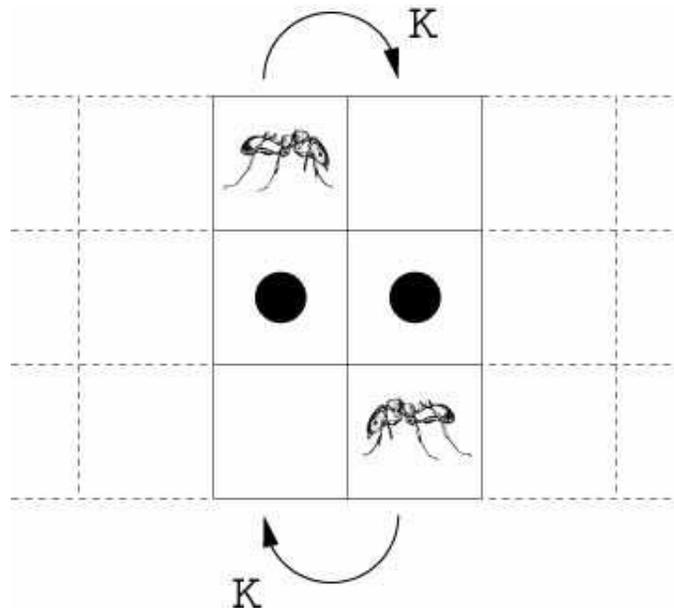}
\end{center}
\caption{A typical head-on encounter of two oppositely moving ants
in the model of {\it bi-directional} ant-traffic \cite{jscn}; the 
corresponding hopping probability is denoted by $K$. This process 
does not have any analog in the model of uni-directional ant-traffic. 
}
\label{fig-modeldef2}
\end{figure}
                                                                                
\begin{figure}[ht]
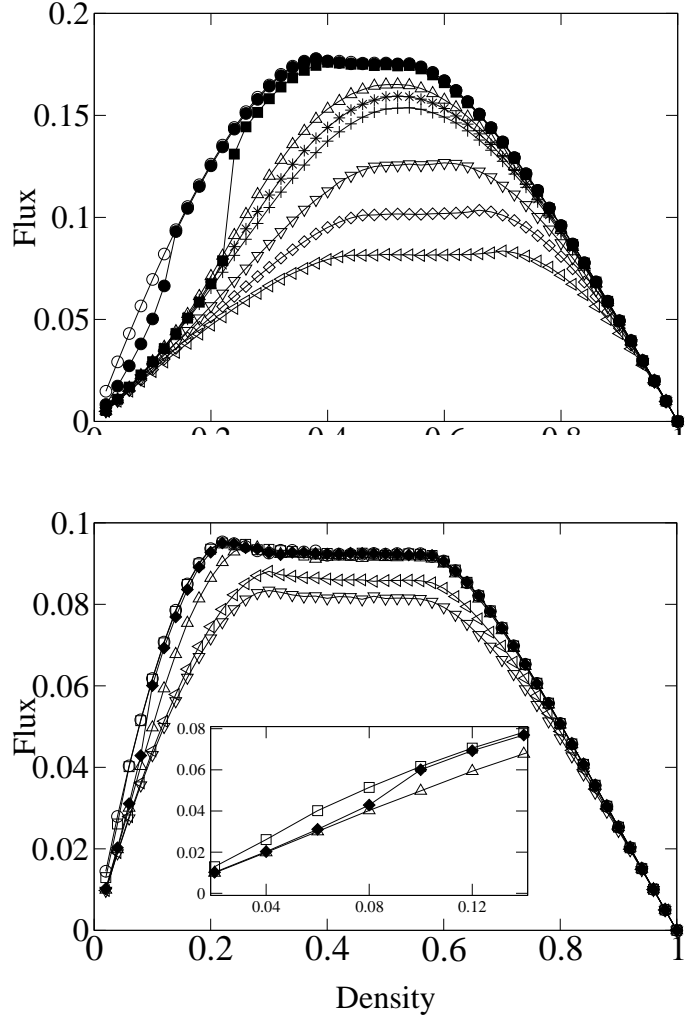

\begin{center}
\ \phantom{a}\\[0.2cm]
\includegraphics[width=0.5\textwidth]{fig28a.eps}
\qquad
\includegraphics[width=0.5\textwidth]{fig28b.eps}
\end{center}
\caption{Fundamental diagrams of the model for bi-directional traffic
\cite{jscn} for the cases $q<K<Q$ (left) and $K<q<Q$ (right) for several 
different values of the pheromone evaporation probability $f$.
The densities for both directions are identical and therefore
only the graphs for one directions are shown. 
The parameters in the left graph are $Q=0.75, q = 0.25$ and $K=0.5$.
The symbols  $\circ$, $\bullet$, $\blacksquare$,
$\bigtriangleup$, $\ast$, $+$, $\bigtriangledown$, $\Diamond$
and $\triangleleft$ correspond, respectively, to
$f = 0, 0.0005, 0.005,0.05, 0.075,0.10,0.25,0.5$ and $1$.
The parameters in the right graph are $Q=0.75, q = 0.50$ and $K=0.25$.
The symbols $\circ$, $\square$, $\blacklozenge$, $\bigtriangleup$,
$\triangleleft$ and $\bigtriangledown$ correspond, respectively, to
$f = 0, 0.0005, 0.005, 0.05, 0.5$ and $1$.
The inset in the right graph is a magnified re-plot of the same data,
over a narrow range of density, to emphasize
the fact that the unusual trend of variation of flux with density in
this case is similar to that observed in the case $q<K<Q$ (left).
The lines are merely guides to the eye. In all cases curves plotted
with filled symbols exhibit non-monotonic behaviour in the speed-density
relation.
}
\label{fig-flux}
\end{figure}

The rules governing the dropping and evaporation of pheromone in the
model of bi-directional ant-traffic are identical to those in the
model of uni-directional traffic. The {\it common} pheromone trail is
created and reinforced by both the outbound and nestbound ants. The
probabilities of forward movement of the ants in the model of
bi-directional ant-traffic are also natural extensions of the similar
situations in the uni-directional traffic. When an ant (in either of
the two lanes) {\it does not} face any other ant approaching it from
the opposite direction the likelihood of its forward movement onto
the ant-free site immediately in front of it is $Q$ or $q$, respectively,
depending on whether or not it finds pheromone ahead. Finally, if an
ant finds another oncoming ant just in front of it, as shown in
Fig.~\ref{fig-modeldef2}, it moves forward onto the next site with probability $K$.
Since ants do not segregate in perfectly well defined lanes, head-on
encounters of oppositely moving individuals occur quite often although
the frequency of such encounters and the lane discipline varies from
one species of ants to another. In reality, two ants approaching each
other feel the hindrance, turn by a small angle to avoid head-on
collision \cite{couzin} and, eventually, pass each other.
At first sight, it may appear that the ants in our model follow perfect
lane discipline and, hence, unrealistic. However, that is not true.
The violation of lane discipline and head-on encounters
of oppositely moving ants is captured, effectively, in an indirect
manner by assuming $K < Q$. But, a left-moving (right-moving) ant
{\it cannot} overtake another left-moving (right-moving) ant immediately
in front of it in the same lane. It is worth mentioning that even
in the limit $K = Q$ the traffic dynamics on the two lanes would
remain coupled because the pheromone dropped by the outbound ants also
influence the nestbound ants and vice versa.
                                                                              
Fig.~\ref{fig-flux} shows fundamental diagrams for the two relevant
cases $q<K<Q$ and $K<q<Q$ and different values of the evaporation
probability $f$ for equal densities on both lanes.
In both cases the unusual behaviour related to
a non-monotonic variation of the average speed with density
as in the uni-directional model can be observed \cite{jscn}.
                                                                               
An additional feature of the fundamental diagram in the bi-directional
ant-traffic model is the occurrence of a plateau region. This plateau
formation is more pronounced in the case $K<q<Q$ than for $q<K<Q$ since
they appear for all values of $f$. Similar plateaus have been observed
earlier \cite{janowsky,tripathy} in models related to vehicular traffic
where randomly placed bottlenecks slow down the traffic in certain
locations along the route.
                                                                               
The experimental data available initially \cite{burd1,burd2} were not 
accurate enough to test the predictions mentioned above. However, 
more accurate recent data \cite{burd3} exhibit non-monotonic variation 
of the average speed with density thereby confirming our theoretical 
prediction.

One of the interesting open questions, which requires careful modelling, 
is as follows: how does a forager ant, which gets displaced from a 
trail, decides the correct direction on rejoining the trail? More 
specifically, an ant carrying food should be nest-bound when it rejoins 
the trail to save time and to minimize the risk of an encounter with a 
predator. In other words, do the pheromone trails have some ``polarity'' 
(analogous to the polarity of microtubules and actin, the filamentary 
tracks on which the cytoskeletal motors move)? On the basis of recent 
experimental observations, it has been claimed \cite{ratnieks1} that 
the trail geometry gives rise to an effective polarity of the ant 
trails. However, other mechanisms for polarity of the trails are also 
possible.

\section{Pedestrian traffic on trails}
\label{sec-ped}


Although there are some superficial similarities between the trafic-like 
collective phenomena in ant-trails and the pedestrian traffic on trails, 
there are also some crucial differences. At present, there are very 
few models which can account for all the observed phenomena in completely 
satisfactory manner.

\subsection{Collective phenomena}
\label{sec-ped-coll}

We present only a brief overview of the collective effects and 
self-organization; for a more comprehensive discussion, see 
ref.\cite{dhrev,PedeProc}. 

{\bf Jamming}: 
At large densities various kinds of jamming phenomena occur, e.g.\
when the flow is limited by a door or narrowing. Therefore, this kind 
of jamming does not depend strongly on the microscopic dynamics 
of the particles, but is typical for a bottleneck situation. It is 
important for practical applications, especially evacuation simulations.
Furthermore, in addition to the flow reduction, effects like arching
\cite{panic,arching}, known from granular materials, play an important 
role. Jamming also occurs where two groups of pedestrians mutually block 
each other. 

\begin{figure}[ht]
  \begin{center}
    a)\quad\includegraphics[width=0.37\textwidth]{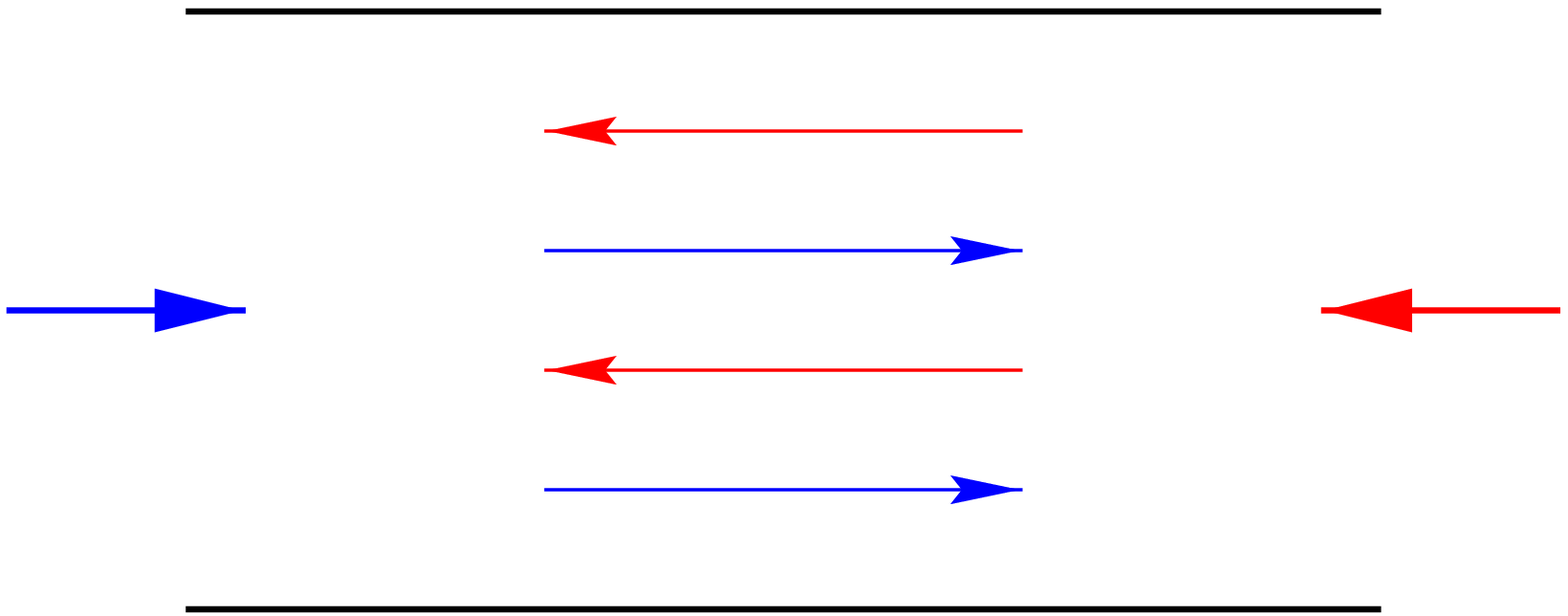}\qquad\qquad
    b)\quad\includegraphics[width=0.43\textwidth]{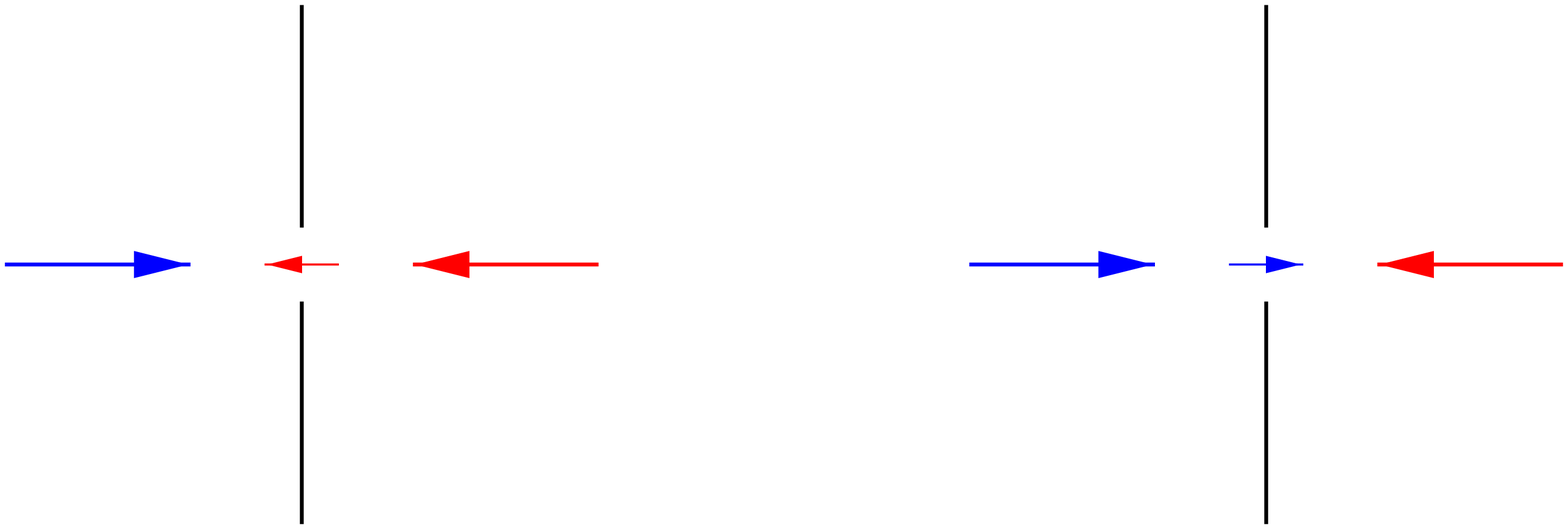}
    \caption{Patterns observed in counterflow. a) Lane formation 
in a narrow corridor. b) Oscillations of the flow direction at a door.}
\label{fig_oszi}
  \end{center}
\end{figure}

{\bf Lane formation}: In counterflow, i.e.\ two groups of people moving 
in opposite directions, a kind of spontaneous symmetry breaking occurs 
(see Fig.~\ref{fig_oszi}a). The motion of the pedestrians can self-organize
in such a way that (dynamically varying) lanes are formed where
people move in just one direction \cite{social}. In this way, strong 
interactions with oncoming pedestrians are reduced and a higher walking 
speed is possible.

{\bf Oscillations}: In counterflow at bottlenecks, e.g.\ doors,
one can observe oscillatory changes of the direction of motion.
Once a pedestrian is able to pass the bottleneck it becomes easier
for others to follow in the same direction until somebody is
able to pass (e.g.\ through a fluctuation) the bottleneck in the
opposite direction (see Fig.~\ref{fig_oszi}b).

{\bf Patterns at intersections}: At intersections various collective
patterns of motion can be formed. Short-lived roundabouts make the 
motion more efficient since they allow for a ``smoother'' motion.

{\bf Panics}: In panic situations, many counter-intuitive phenomena 
can occur. In the faster-is-slower effect \cite{panic} a higher 
desired velocity leads to a slower movement of a large crowd. 
Understanding such effects is extremely important for evacuations in 
emergency situations.

\subsection{Modelling Pedestrian Dynamics}

Several different approaches for modelling the dynamics of pedestrians
have been proposed, either based on a continuous representation of 
space or on a grid. 

The earliest models of pedestrian dynamics belonged to the population-based
approaches and took inspiration from hydrodynamics or gas-kinetic theory
\cite{hydro,hhydro,hughes1,hughes2}. However, it turned our \cite{hhydro}
that several important differences to normal fluids are important,
e.g. the anisotropy of interactions or the fact that pedestrians usually
have an individual preferred direction of motion.
                                                                                
Later several individual-based approaches in continuous space and time
have been proposed. In the social force models (see e.g.\ \cite{dhrev,social}
and references therein) and other similar approaches \cite{TM95} pedestrians
are treated as particles subject to repulsive forces induced by the social
behaviour of the individuals. This leads to (coupled) equations of motion
similar to Newtonian mechanics. There are, however, important differences
since, e.g., in general the third law (``actio = reactio'') is not fulfilled.
Furthermore a two-dimensional variant of the optimal-velocity model \cite{OV}
has also been suggested.

Active walker models \cite{activewalker,trail} have been
used to describe the formation of human or animal trails etc.
Here the walker leaves a trace by modifying the underground on his path.
This modification is real in the sense that it could be measured in
principle. For trail formation, vegetation is destroyed by the walker.
                                                                               
In \cite{chopard2} a kind of mesoscopic approach inspired by
lattice gas models \cite{droz} has been suggested. Thus the
exclusion principle is relaxed and the dynamics is based on
a collision-propagation scheme.
                                                                               
Most cellular automaton models for pedestrian dynamics proposed are
rather simple \cite{fukui,nagatani,hubert} and can be considered as
two-dimensional generalizations of the ASEP (see Sec.~\ref{sec-asep}).
However, these models are not able to reproduce all the collective
effects described in the preceeding subsection. The same is true for more
sophisticated discrete models \cite{gipps,bolay}.
                                                                               
In the following we discuss a promising new approach, the {\em floor field
model} \cite{ourpaper,ourpaper2,friction,discrete}, for the description
of pedestrian dynamics in more detail which is takes inspiration from the
ant trail model of Sec.~\ref{sec-ants}.
The interaction in this model is implemented as 'virtual chemotaxis'
which allows to translate a long-ranged spatial interaction into a
local interaction with ``memory''.
Guided by the phenomenon of chemotaxis the interactions between
pedestrians are thus local and allow for computational efficiency.


\subsection{Floor field cellular automaton model}

Pedestrians are modelled as particles that move on a two-dimensional
lattice of cells. Each cell can be occupied by at most one particle
which reflects that the interactions between them are repulsive for
short distances ('private sphere').
Particles can move to one of the neighbouring cells based on certain
transition probabilities that are are determined by three factors:
(1) the desired direction of motion,
(2) interactions with other pedestrians,  and (3) interactions with
the infrastructure (walls, doors, etc.).
                                                                               
First of all, basic transition probabilities are determined which
reflect the preferred walking direction and speed of each
individual in the form of a matrix of preferences $M$ which
can be related to the preferred velocity vector and its fluctuations
\cite{ourpaper}.
                                                                               
Next interactions between pedestrians are taken into account. The exclusion
principle accounts for the fact that one likes to keep a minimal distance 
from others. However, for larger distances the interaction is assumed to 
be attractive to capture the advantage in following the predecessor.
This is implemented by virtual chemotaxis. Moving particles
create a "pheromone" at the cell which they leave, thus creating a kind
of trace, the {\em dynamic floor field} $D$. It has its own dynamics given
by diffusion and decay \cite{ourpaper} which leads to a dilution and
finally the vanishing of the trace after some time.
                                                                               
For a unified description another floor field is introduced, the {\em
static floor field} $S$. It is constant and takes into account the
interactions with the infrastructure, e.g.\ preferred areas, walls and
other obstacles.

The transition probabilities given by the matrix of preference $M_{ij}$
are now modified by the strengths of the floor fields $D_{ij}$ and $S_{ij}$
in the target cell $(i,j)$; the details are given in ref.\cite{ourpaper}. 
Due to the use of parallel dynamics it might happen that two (or more) 
pedestrians choose the same target cell. Such a situation is called a 
{\em conflict}. Due to hard-core exclusion, at most one persons can move.
Introducing a friction parameter $\mu$ \cite{friction}, with probability 
$\mu$ {\em all} pedestrians remain at their site, i.e.  nobody is allowed 
to move. With probability $1-\mu$ one of the individuals is chosen randomly 
and allowed to move to the target cell. The effects of $\mu>0$ are similar 
to those arising from a moment of hesitation when people are about to 
collide and try to avoid the collision.

The details of the update rules can be found in \cite{ourpaper,friction}; 
this floor field model has been able to reproduce the empirically observed 
phenomena listed earlier in this section. Another interesting phenomenon 
which has been studied successfully using the floor field model is the 
evacuation from a large space with only one exit \cite{ourpaper2,friction}.
However, this is beyond the scope of this review as we restrict our 
attention here mostly on traffic-like flow properties.
Nevertheless the application of such models in the planning stages
of large buildings, ships or aircrafts has become increasingly
important over the last years. The simplicity and realism of the
models allows to optimize evacuation and egress processes already
in an early stage of the construction without the necessity of
performing potentially dangerous experiments.

\section{Summary and conclusion}
\label{sec-sum}

Because of restrictions imposed by the allowed length of the review, we 
have excluded several biological traffic phenomena where, to our knowledge, 
very little progress has been made so far in theoretical modelling of the 
processes. These include, for example, \\
(i) {\it bidirectional transport} along microtubules where the same cargo 
moves along the same microtubule track using sets of opposing motors 
\cite{welte,gross};\\ 
(ii) {\it self-organized patterns} like, for example, {\it asters} and 
{\it vortices}, which have been in several in-vitro experiments 
\cite{nedlec} and model calculations \cite{kardar,sunil,zimmer}

In this article we have reviewed our current understanding of traffic-like 
collective phenomena in living systems starting from the smallest level 
of intra-cellular transport and ending at the largest level of traffic of 
pedestrians. 
So far as the theoretical methods are concerned, we have restricted our 
attention to those works where the language of cellular automata or 
extensions of TASEP has been used. The success of this modelling 
strategy has opened up a new horizon and, we hope, we have provided a 
glimpse of the exciting frontier.

\vspace{1cm}

\noindent {\bf Acknowledgements:} It is our great pleasure to thank 
Yasushi Okada, Alexander John and Ambarish Kunwar for enjoyable 
collaborations on the topics discussed in this review. We are indebted 
to many colleagues for illuminating discussions although we are unable 
to mention all the names because of space limitations.


\end{document}